\documentclass[onecolumn]{aastex62}  

\newcommand\apjcls{1}
\newcommand\aastexcls{2}
\newcommand\othercls{3}

\newcommand\papercls{\aastexcls}

\if\papercls \apjcls
\usepackage{apjfonts}
\else\if\papercls \othercls
\usepackage{epsfig}
\usepackage{margin}
\usepackage{times}
\fi\fi
\usepackage{ifthen}
\usepackage{natbib}
\usepackage{amssymb, amsmath}
\usepackage{appendix}
\usepackage{etoolbox}
\usepackage[T1]{fontenc}
\usepackage{paralist}

\if\papercls \apjcls
\newcommand\aas{\ref@jnl{AAS Meeting Abstracts}}
\newcommand\dps{\ref@jnl{AAS/DPS Meeting Abstracts}}
\newcommand\maps{\ref@jnl{MAPS}}
\else\if\papercls \othercls
\usepackage{astjnlabbrev-jh}
\fi\fi

\if\papercls \aastexcls
\hypersetup{citecolor=blue, 
            linkcolor=blue, 
            menucolor=blue, 
            urlcolor=blue}  
\else
\usepackage[
bookmarks=true,           
bookmarksnumbered=true,   
colorlinks=true,          
citecolor=blue,           
linkcolor=blue,           
menucolor=blue,           
urlcolor=blue,            
linkbordercolor={0 0 1},  
pdfborder={0 0 1},
frenchlinks=true]{hyperref}
\fi

\if\papercls \othercls

\else

\fi

\providecommand{\adsurl}[1]{\href{#1}{ADS}}

\makeatletter
\patchcmd{\NAT@citex}
  {\@citea\NAT@hyper@{%
     \NAT@nmfmt{\NAT@nm}%
     \hyper@natlinkbreak{\NAT@aysep\NAT@spacechar}{\@citeb\@extra@b@citeb}%
     \NAT@date}}
  {\@citea\NAT@nmfmt{\NAT@nm}%
   \NAT@aysep\NAT@spacechar\NAT@hyper@{\NAT@date}}{}{}

\patchcmd{\NAT@citex}
  {\@citea\NAT@hyper@{%
     \NAT@nmfmt{\NAT@nm}%
     \hyper@natlinkbreak{\NAT@spacechar\NAT@@open\if*#1*\else#1\NAT@spacechar\fi}%
       {\@citeb\@extra@b@citeb}%
     \NAT@date}}
  {\@citea\NAT@nmfmt{\NAT@nm}%
   \NAT@spacechar\NAT@@open\if*#1*\else#1\NAT@spacechar\fi\NAT@hyper@{\NAT@date}}
  {}{}
\makeatother

\makeatletter
\DeclareRobustCommand{\lowcase}[1]{\@lowcase#1\@nil}
\def\@lowcase#1\@nil{\if\relax#1\relax\else\MakeLowercase{#1}\fi}
\makeatother

\DeclareSymbolFont{UPM}{U}{eur}{m}{n}
\DeclareMathSymbol{\umu}{0}{UPM}{"16}
\let\oldumu=\umu
\renewcommand\umu{\ifmmode\oldumu\else\math{\oldumu}\fi}

\if\papercls \othercls

\else

\fi

\let\oldsim=\sim
\renewcommand\sim{\ifmmode\oldsim\else\math{\oldsim}\fi}
\let\oldpm=\pm
\renewcommand\pm{\ifmmode\oldpm\else\math{\oldpm}\fi}
\newcommand\by{\ifmmode\times\else\math{\times}\fi}

\newbox{\wdbox}
\renewcommand\c{\setbox\wdbox=\hbox{,}\hspace{\wd\wdbox}}
\renewcommand\i{\setbox\wdbox=\hbox{i}\hspace{\wd\wdbox}}

\newcount\timect
\newcount\hourct
\newcount\minct
\newcommand\now{\timect=\time \divide\timect by 60
         \hourct=\timect \multiply\hourct by 60
         \minct=\time \advance\minct by -\hourct
         \number\timect:\ifnum \minct < 10 0\fi\number\minct}

\catcode`@=11

\newcommand\comment[1]{}

\newcommand\commenton{\catcode`\%=14}

\renewcommand\math[1]{$#1$}
\newcommand\mathshifton{\catcode`\$=3}

\let\atab=&
\newcommand\atabon{\catcode`\&=4}

\let\oldmsp=\sp
\let\oldmsb=\sb
\def\sp#1{\ifmmode
           \oldmsp{#1}%
         \else\strut\raise.85ex\hbox{\scriptsize #1}\fi}
\def\sb#1{\ifmmode
           \oldmsb{#1}%
         \else\strut\raise-.54ex\hbox{\scriptsize #1}\fi}
\newbox\@sp
\newbox\@sb
\def\sbp#1#2{\ifmmode%
           \oldmsb{#1}\oldmsp{#2}%
         \else
           \setbox\@sb=\hbox{\sb{#1}}%
           \setbox\@sp=\hbox{\sp{#2}}%
           \rlap{\copy\@sb}\copy\@sp
           \ifdim \wd\@sb >\wd\@sp
             \hskip -\wd\@sp \hskip \wd\@sb
           \fi
        \fi}
\def\msp#1{\ifmmode
           \oldmsp{#1}
         \else \math{\oldmsp{#1}}\fi}
\def\msb#1{\ifmmode
           \oldmsb{#1}
         \else \math{\oldmsb{#1}}\fi}

\def\supon{\catcode`\^=7}

\def\subon{\catcode`\_=8}

\def\supsubon{\supon \subon}

\newcommand\actcharon{\catcode`\~=13}

\newcommand\paramon{\catcode`\#=6}

\comment{And now to turn us totally on and off...}

\newcommand\reservedcharson{ \commenton  \mathshifton  \atabon  \supsubon 
                             \actcharon  \paramon}

\catcode`@=12
\reservedcharson

\if\papercls \apjcls

\else

\fi

\newcommand\chisq{\ifmmode{\chi\sp{2}}\else\math{\chi\sp{2}}\fi}
\newcommand\redchisq{\ifmmode{ \chi\sp{2}\sb{\rm red}}
                    \else\math{\chi\sp{2}\sb{\rm red}}\fi}
\newcommand\Teq{\ifmmode{T\sb{\rm eq}}\else$T$\sb{eq}\fi}
\newcommand\Teff{\ifmmode{T\sb{\rm eff}}\else$T$\sb{eff}\fi}
\newcommand\mjup{\ifmmode{M\sb{\rm Jup}}\else$M$\sb{Jup}\fi}
\newcommand\rjup{\ifmmode{R\sb{\rm Jup}}\else$R$\sb{Jup}\fi}
\newcommand\msun{\ifmmode{M\sb{\odot}}\else$M\sb{\odot}$\fi}
\newcommand\rsun{\ifmmode{R\sb{\odot}}\else$R\sb{\odot}$\fi}
\newcommand\mearth{\ifmmode{M\sb{\oplus}}\else$M\sb{\oplus}$\fi}
\newcommand\rearth{\ifmmode{R\sb{\oplus}}\else$R\sb{\oplus}$\fi}

\newcommand\degree{\degr}

\usepackage{amsmath}
\usepackage{graphicx}

\usepackage{epsfig}
\usepackage{natbib}
\usepackage{wasysym}
\usepackage{lipsum}
\usepackage{mathrsfs}
\usepackage{float}
\usepackage{rotating}
\usepackage{graphicx}
\usepackage{epstopdf}
\usepackage{color}

\definecolor{twitterblue}{RGB}{64,153,255}

\usepackage{hyperref}

\usepackage{filecontents}

\begin{document}

\shorttitle{Influence of Stellar Flares on Atm. Chemistry}
\shortauthors{Chen, H.,  Zhuchang, Z., Youngblood, A., Wolf, E.T., Feinstein, A.D., Horton, D.E.}

\title{{\bf Persistence of Flare Driven Atmospheric Chemistry on Rocky Habitable Zone Worlds}}

\author[0000-0003-1995-1351]{Howard Chen}
\affil{ Center for Interdisciplinary Exploration \& Research in Astrophysics (CIERA), Evanston, IL 60202, USA}
\affil{Department of Earth and Planetary Sciences, Northwestern University, Evanston, IL 60202}

\author[0000-0002-4142-1800]{Zhuchang Zhan}
\affil{ Department of Earth, Atmospheric, and Planetary Sciences, Massachusetts Institute of Technology, Cambridge, MA 02138, USA}

\author[0000-0002-1176-3391]{Allison Youngblood}

\affil{Laboratory for Atmospheric and Space Physics, University of Colorado Boulder, Boulder, CO 80309, USA}

\author[0000-0002-7188-1648]{Eric T. Wolf}

\affil{Laboratory for Atmospheric and Space Physics, University of Colorado Boulder, Boulder, CO 80309, USA}
\affil{NASA Nexus for Exoplanet System Science Virtual Planetary Laboratory, Seattle, WA 98194, USA}
\affil{NASA Goddard Sellers Exoplanet Environments Collaboration, Greenbelt, MD,  USA}

\author[0000-0002-9464-8101]{Adina D. Feinstein}

\affil{ Department of Astronomy \& Astrophysics, University of Chicago, Chicago, IL 60637, USA }

\author[0000-0002-2065-4517]{Daniel E. Horton}

\affil{ Center for Interdisciplinary Exploration \& Research in Astrophysics (CIERA), Evanston, IL 60202, USA}
\affil{Department of Earth and Planetary Sciences, Northwestern University, Evanston, IL 60202}

\correspondingauthor{Howard Chen} \email{howard@earth.northwestern.edu}

\begin{abstract}
Low-mass stars show evidence of vigorous magnetic activity in the form of large flares and coronal mass ejections. Such space weather events may have important ramifications for the habitability and observational fingerprints of exoplanetary atmospheres. Here, using a suite of three-dimensional coupled chemistry-climate model (CCM) simulations, we explore effects of time-dependent stellar activity on rocky planet atmospheres orbiting G-, K-, and M-dwarf stars. We employ observed data from the MUSCLES campaign and {\it Transiting Exoplanet Satellite Survey} and test a range of rotation period, magnetic field strength, and flare frequency assumptions. We find that recurring flares drive K- and M-dwarf planet atmospheres into chemical equilibria that substantially deviate from their pre-flare regimes, whereas G-dwarf planet atmospheres quickly return to their baseline states. Interestingly, simulated O$_2$-poor and O$_2$-rich atmospheres experiencing flares produce similar mesospheric nitric oxide abundances, suggesting that stellar flares can highlight otherwise undetectable chemical species. Applying a radiative transfer model to our CCM results, we find that flare-driven transmission features of bio-indicating species, such as nitrogen dioxide, nitrous oxide, and nitric acid, show particular promise for detection by future instruments.
\end{abstract}

\section*{Introduction} 
\label{sec:intro}
In recent years, large-scale observational campaigns such as the {\it Kepler Space Telescope} \citep{BoruckiEt2010SCI} indicate that rocky planets are common \citep{KopparapuEt2013ApJL,MuldersEt2018AJ,HsuEt2019AJ,BrysonEt2020arXiv}. A handful of these systems are known to reside within the circumstellar habitable zones (HZs) of their host stars \citep{KastingEt1993Icarus,KastingEt2015ApJL} and  are amenable to atmospheric spectroscopic measurements (for example, Kepler-186, Kepler-452, Proxima Centauri, TRAPPIST-1, LHS-1140, and TOI-700). On-going Transiting Exoplanet Satellite Survey (TESS) operations \citep{RickerEt14} will discover many closer and brighter planetary systems, offering more terrestrial exoplanets for follow-up mass measurements and atmospheric characterization efforts \citep{Ballard2019AJ,DalbaEt2019PASP}. However, interpretation of remotely sensed atmospheric data will require making sense of a planet’s signals in context, i.e., understanding interactions of atmospheric chemistry, physics, dynamics, and thermodynamics, within the space weather environment and electromagnetic radiation regime. For instance, strong stellar chromospheric activity from low-mass stars could influence attendant atmospheres via the dissociation, excitation, and ionization processes associated with space weather events \citep{ScaloEt2007AsBio,LinkskyEt2017A&A} -- leading to substantial alteration of planetary atmospheres and chemical signatures \citep{AirapetianEt2019AsBio}. Modulation of atmospheric chemistry can have a critical influence on the bulk atmospheric composition \citep{Tian&Ida2015NatGeo,BeckerEt2020AJ}, atmospheric dynamics \citep{ThurairajahEt2019JGR}, surface radiation dosage \citep{Atri2017MNRAS,YamashikiEt2019ApJ}, and detectability of atmospheric species \citep{AirapetianEt2017Sci}.

Stellar activity -- which includes stellar flares, coronal mass ejections (CMEs), and stellar proton events (SPEs) --  has a profound influence on a planet’s habitability, primarily via its influence on atmospheric ozone. Stellar flares are rapid (minutes to days) releases of coronal magnetic energy accompanied by bursts of electromagnetic radiation and accelerated ionized particle fluences. Previous work has shown that while a single large flare (e.g., AD Leonis Great Flare with UV-optical energy of $E \sim 10^{34}$ erg; \citep{Hawley+Petterson1991ApJ}) does not substantially affect an Earth-like planet’s habitability \citep{SeguraEt2010AsBio}, repeated secular flaring and CMEs, by way of UV emissions ($\lambda < 350$ nm) and SPEs ($E > 100$ MeV), could destroy a planet's ozone layer within a few Earth-years \citep{TilleyEt2019AsBio}. With the exception of the strongest flaring scenarios, time-independent models have predicted that initially Earth-like atmospheres should retain appreciable amounts of ozone that could efficiently filter out incident UV-B and UV-C radiation \citep{GrenfellEt2012AsBio}. However, this conclusion stands in contrast with that of time-resolved models, in which nearly 90\% of the stratospheric ozone column could be eroded, even with conservative assumptions in flare magnitude and frequency \citep{TilleyEt2019AsBio}.

In addition to stellar flares' considerable influence on ozone, flares are likely to have observationally-relevant  effects on planetary atmospheres \citep{VenotEt2016ApJ}. Remote detection of biological processes, or equivalently, the measure of thermodynamic chemical disequilibrium \citep{Krissansen-TottonEt2018SCI,SchweitermanEt2018AsBio}, is linked to atmospheric chemistry, brightness temperature, and stellar variability. In the search for signals of life, detection of either biosignatures, i.e., atmospheric constituents that are the direct result of life processes \citep{DesMaraisEt2002AsBio,SchweitermanEt2018AsBio}, or bio-indicators, i.e., the photochemical derivatives of biosignatures, is sought. High magnitude stellar flares may trigger disequilibrium chemistry and generate bio-indicating species through photo-excitation, photo-dissociation, and/or strong mixing. For example, it has been hypothesized that enhanced accumulation of nitric acid (HNO3) in a biologically active atmosphere experiencing flares could serve as a bio-indicator, as purely abiotic concentrations of HNO3 in anoxic conditions should be low \citep{TabatabaEt2016A&A,ScbeucherEt2018ApJ}. Abiotic nitric oxide (NO) and nitrogen dioxide (NO$_2$) production in O$_2$-depleted environments is also not expected to result in substantial atmospheric accumulation as their production is mediated by collisions between molecular biogenic oxygen and nitrogen. In atmospheres dominated by N$_2$–O$_2$–H$_2$O however, stellar activity could highlight species such as NO, OH, O$_2$($^1$D) and CO$_2$ to potentially detectable levels \citep{AirapetianEt2017Sci,TilleyEt2019AsBio}.

Previous efforts to understand the effects of stellar flares on planetary atmospheres underscore the value of single column climate models. Their efficiency and speed allow for a wide sampling of the planetary parameter space. However, lower-dimensional models do not account for critical three-dimensional (3D) processes, including atmospheric circulation, large-scale mixing, and cloud dynamics -- factors that have substantial impact on the climate and chemistry of exoplanets, particularly for slow-rotators orbiting late K-dwarf and early M-dwarf stars \citep{YangEt2013ApJL,ChenEt2018ApJL,ChenEt2019ApJ}. Assessments of stellar chromospheric and coronal activity influences on planetary atmospheres are likewise inherently 3D problems, e.g., the role of a magnetic field, inclusion of dynamical mixing, and illumination geometry. As such, 3D considerations are likely to have crucial consequences on flare influence predictions. Furthermore, the majority of previous exoplanet-flare-habitability studies have used UV observations from a single high magnitude $10^{34}$ erg superflare (AD Leonis; \citep{Hawley+Petterson1991ApJ,Atri2017MNRAS}), that is unlikely to be representative of typical stellar flaring behavior. Incorporating UV flare spectra and lightcurve temporal evolution using more typical flaring behavior may offer new insights into the photochemical characteristics of habitable zone planets \citep{LoydEt2018ApJa,PeacockEt2019ApJ}.

To better understand interactions between flaring stars and their attendant planets, in this study we employ a 3D global Earth System CCM, the Whole Atmosphere Community Climate Model (WACCM; \cite{MarshEt2013JGR}). We simulate nitrogen-dominated rocky exoplanet atmospheres experiencing time-dependent stellar UV activity and proton events around G-, K-, and M-dwarf stars (Extended Data Figure 1-3). Due to i) stationary substellar cloud formation that prevents runaway greenhouse states near the inner edge of the habitable zone \citep{YangEt2013ApJL,WayEt2016GRL}, ii) decreased ice-albedo feedbacks which inhibit total glaciation near the outer edge of the habitable zone \citep{ShieldsEt2014ApJ,CheclairEt2017ApJ,CheclairEt2019ApJL},  and iii) increased nutrient replenishment via oceanic upwelling \citep{OlsonEt2020ApJ}, previous 3D studies have demonstrated that slowly- and synchronously-rotating ocean-covered planets have increased habitability potential. With the inclusion of time-dependent stellar flares, the results from this work further the notion that slowly- and synchronously-rotating planets are favorable targets by future instruments.

\newpage

\begin{figure*}[h] 
\begin{center}
\includegraphics[width=.7\columnwidth]{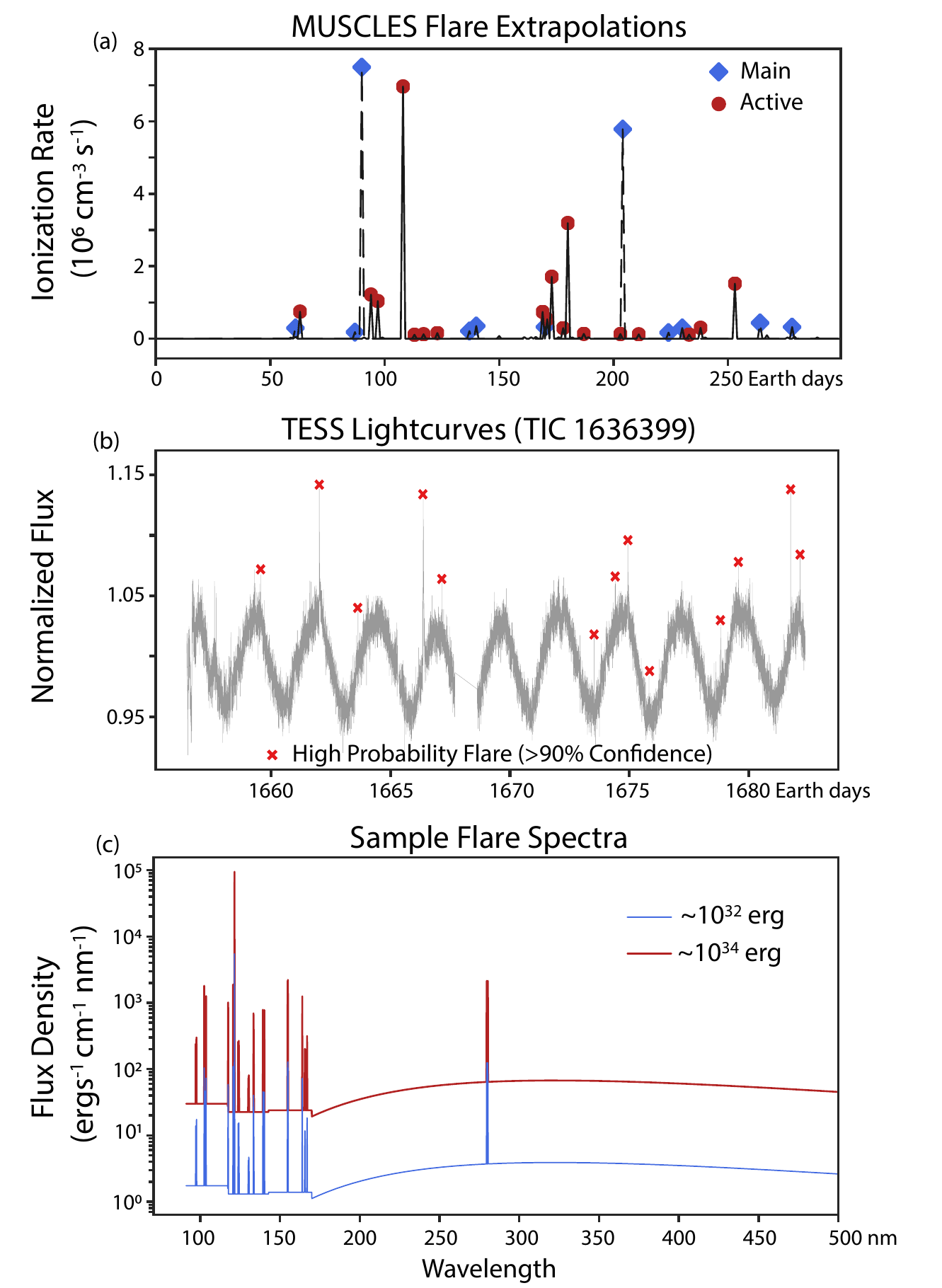}
\caption{\label{fig:input}  {\bf Observed flare lightcurves and spectra used as inputs for CCM simulations.} {\bf a,} Flare-driven atmospheric ionization rates produced via power-law extrapolations for large flares ($E > 10^{32}$ erg) from the MUSCLES survey. Data that underpins the main text are denoted by ``main”, whereas data approaching the activity level of Proxima Centauri are denoted by ``active” (See Methods). The dashed black lines
connect the ‘main’ input ionization rates and the solid black line connect those of
the ‘active’. {\bf b,} Observed TESS lightcurves from TIC 1636399 identified by convolution neural network. {\bf c,} Sample MUSCLES spectra used in model simulations for flares with two different approximate energies during their impulsive phases.       } 
\end{center}
\end{figure*}

\section*{Numerical Setup}

We perform simulations of rocky exoplanetary atmospheres orbiting G-, K-, and M-star archetypes (Sun-like star, HD85512, and TRAPPIST-1). For each planet scenario, we test different oxygenation states, magnetosphere strengths, and stellar activity levels (Supplementary Table 1 and Extended Data Figure 2-3). Pre-flare baseline atmospheric compositions are: N$_2$ (78\%), CH$_4$ (0.701 ppmv), N$_2$O (0.273 ppmv), and CO$_2$ (288 ppmv). We focus on three different scenarios: (1) a magnetized rapidly rotating planet around a Sun-like star (which we refer to as ``G-star planet”), (2) an unmagnetized slow rotator around HD85512 (``K-star planet”), and (3) a weakly magnetized rapid rotator around TRAPPIST-1 (``M-star planet”). The latter two scenarios assume synchronous-rotation. For each scenario, we simulate the effects of O$_2$-rich (modern-Earth-like) and O$_2$-poor (Proterozoic-Earth-like) initial atmospheres. Configurations and boundary conditions were chosen for self-consistency, that is, simulated configurations follow from known physics.  For example,  greater orbital separations  of synchronously-rotating planets around early M-dwarfs imply slower rotation (compared to those around late M-dwarfs). Without rotation induced convection of conducting inner core fluids, i.e., a magnetic dynamo, these planets are unlikely to sustain strong planetary-scale magnetic fields \citep{ChristensenEt2009NAT}. Thus, while we test the sensitivity of our results to a suite of input parameters and assumptions, the main text presents the most self-consistent simulation scenarios, corresponding to experiments 1, 4, and 8 in Supplementary Tables 1 and 2.

 Steady-state stellar spectral energy distributions  from the 1850 Solar Irradiance spectrum \citep{lean1995reconstruction}, TRAPPIST-1 (M8V; $T_{\rm eff}$ = 2511 K; Wilson et al. in review), and HD 85512 (K6V; $T_{\rm eff}$ = 4715 K; version 2.2; \cite{FranceEt2016ApJ} are used as inputs to the CCM. To compute time-dependent activity (Figure 1), we utilize a Measurements of the Ultraviolet Spectral Characteristics of Low-mass Exoplanetary Systems (MUSCLES) flare generator \citep{LoydEt2018ApJa} and observed {\it TESS} flare data identified by a convolutional neural network \citep{FeinsteinEt2020arXiv}. Proton fluence calculations are derived from the above electromagnetic flares \citep{YoungbloodEt2017ApJ} and all CCM experiments described are subject to the same stellar activity time series inputs (see Methods and Extended Data Figure 2-3). 

We first report the effects of MUSCLES-derived stellar flares on our self-consistent scenarios. We then explore the effects of observed {\it TESS} flares, discuss the coupled effects of atmospheric transport, and present results for key flare-modulated chemical species. Next, we compare and contrast flare-induced differences between O$_2$-rich versus O$_2$-poor atmospheres and moist greenhouse states versus temperate states.  Finally, we discuss the observational
implications of our results. See the Supplementary Information for the complete simulation setups and input data.

\begin{figure*}[h] 
\begin{center}
\includegraphics[width=0.75\columnwidth]{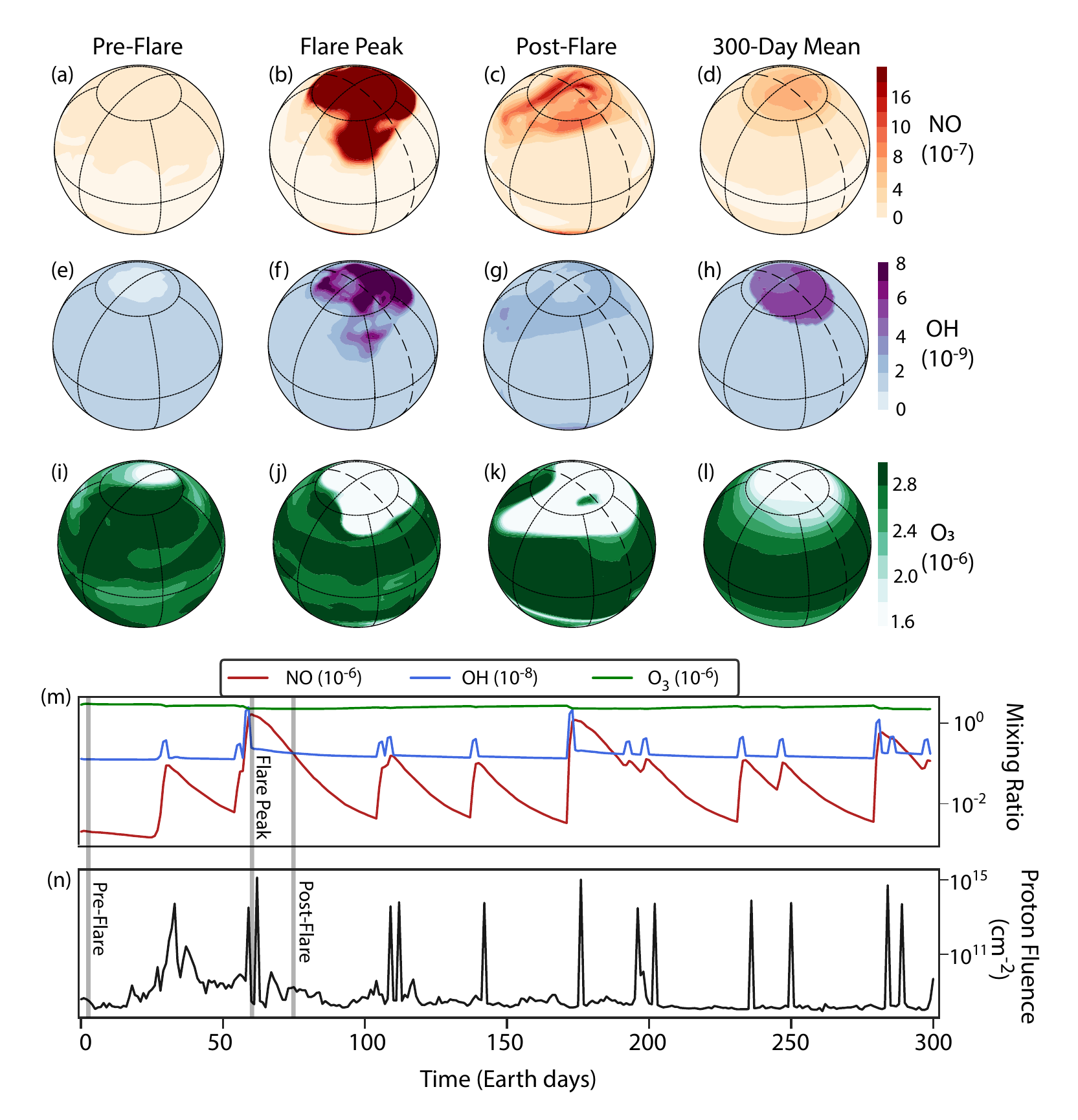}
\caption{\label{fig:bts1}  {\bf Spatial and temporal atmospheric effects of repeated stellar flares on a G-star planet.} Simulated  global time slice distributions of upper atmospheric NO (a-d), OH (e-h), and O$_3$ (i-l) mixing ratios  and their global average time-series (m) that result from exposure to time evolving flare-derived proton fluences (n). The simulated planet rotates around a Sun-like star non-synchronously and has a magnetic field. NO and OH mixing ratios are reported at 0.1 hPa, whereas O$_3$ mixing ratios are reported at 1.0 hPa.  Spherical projections are centered on $40\degree$ N latitude and $225\degree$ longitude.  } 
\end{center}
\end{figure*} 

\section*{3D Effects of Large Stellar Flares (Results)}

We find that strong stellar flares drive dramatic transient and steady-state changes in stratospheric and mesospheric chemistry, particularly in nitrogen and hydrogen oxide reservoirs. Our results show that energetic flares ($E > 10^{33}$ ergs and proton fluence $> 10^{14}$ cm$^{-2}$) have profound impacts on atmospheric species such as nitric oxide (NO), hydroxide (OH), and ozone (O$_3$) (Figures 2, 3, and Extended Data Figure 4; Supplementary Table 1 Exp. 1, 4, and 8). On unmagnetized K- and M-star planets in particular, modeled stellar flares modulate the atmospheric concentration of many photochemically important species and ultimately establish new chemical steady-state regimes that substantially deviate from their pre-flare compositions (Figure 3m and Extended Data Figure 4m). In contrast, flares do not substantially perturb the atmospheric compositions of magnetized G-star and TESS planets (Figure 2m; Extended Data Fig. 5).

Simulated mesospheric nitrogen oxides such as NO, derived from reactions initiated by precipitating electrons \citep{jackman2005neutral}, are orders of magnitude more abundant on K- and M-star planets (Figures 3a-d and Extended Data Figure 4a-d; Supplementary Table 1 Exp. 4 and 8) than those around G-stars (Figure 2a-d; Supplementary Table 1 Exp. 1). This is the combined result of the greater latitudinal extents of proton deposition for weakly or unmagnetized planets (see Methods), steady-state inputs of UV-B spectra, illumination geometries, and planetary circulation regimes.

The circumstellar UV photochemical environment and slow rotation of K- and M-star planets, leads to the persistence of mesospheric NO. NO mixing ratios on our simulated K- and M-star planets do not return to their pre-flare levels after a large flare (Figure 3m), whereas NO on G-star planets returns to pre-flare levels within $<50$ days (Figure 2m). Enhanced simulated global-mean NO lifetimes on M-star planets are due to lesser emission of UV-B radiation ($280 < \lambda < 315$ nm) by their host stars, which promotes O($^1$D) formation, reducing an OH sink (as H$_2$O$_v$ + O($^1$D) = 2OH). In addition, prolonged NO$_{\rm x}$ lifetimes on slowly-rotating K-star planets are likely due to thermally driven radial day-to-night advection that transports produced substellar NO$_{\rm x}$ to the nightside,  where it is temporarily stored. Here, daytime nitrogen-hydrogen oxide chemistry, such as NO + HO$_2$ and NO + NO$_2$ photolysis,  is averted (see Methods), leading to the time averaged enhancement of NO$_{\rm x}$ abundances. In contrast, rapid horizontal mixing and higher incident UV-B radiation for planets around G-stars ($P = 24$ hr) leads to more efficient NO$_{\rm x}$ removal via reaction with ozone and direct titration (Figure 2a-d).

\begin{figure*}[t] 
\begin{center}
\includegraphics[width=0.75\columnwidth]{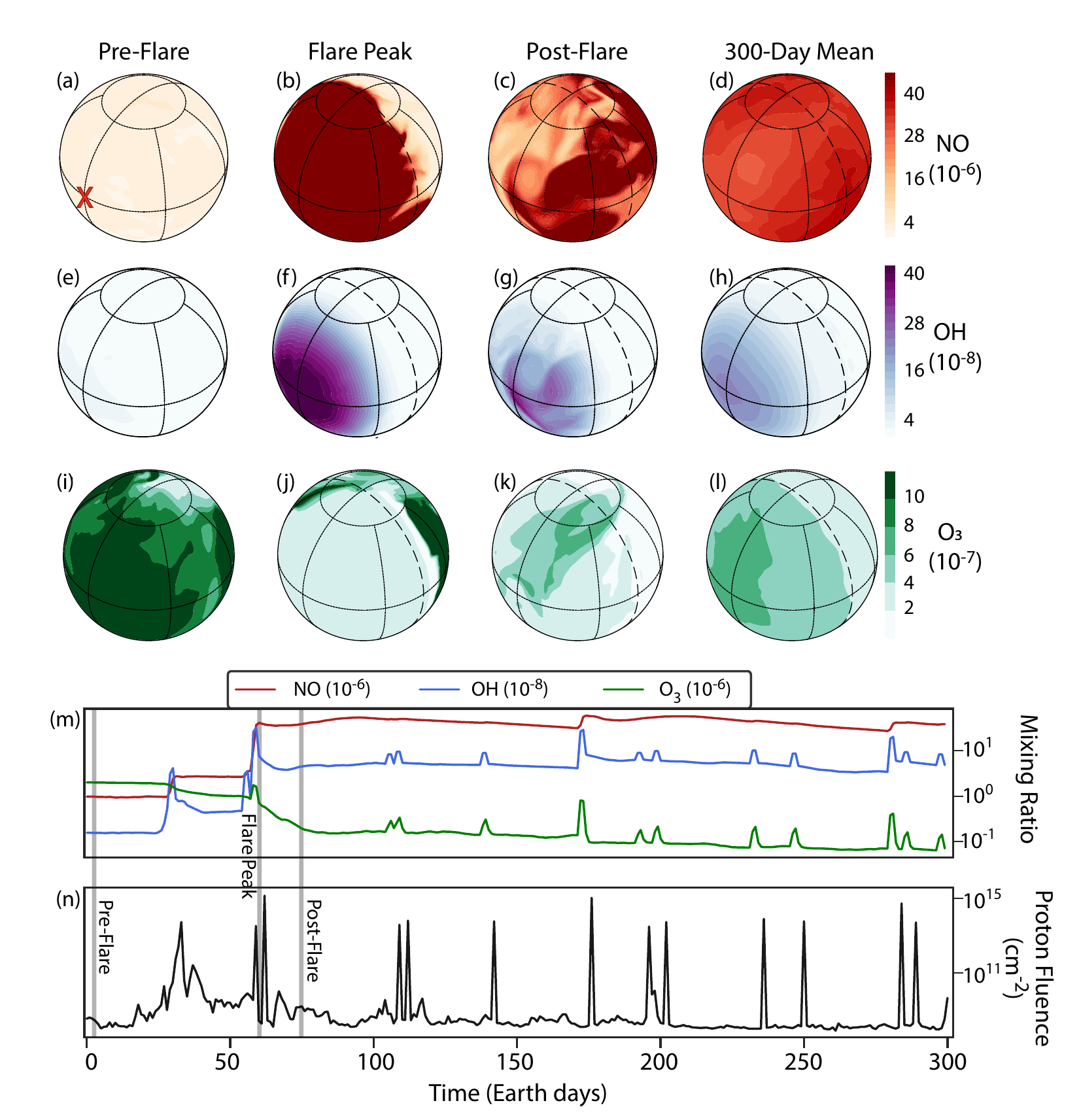}
\caption{\label{fig:bts2} {\bf Spatial and temporal atmospheric effects of repeated stellar flaring on K-star planet.}  Simulated  global time slice distributions of upper atmospheric NO (a-d), OH (e-h), and O$_3$ (i-l) mixing ratios   and their global average time-series (m) that result from exposure to flares with time-evolving proton fluences (n). The simulated planet rotates around K-star HD85512 synchronously and does not have a magnetic field.  NO and OH mixing ratios are reported at 0.1 hPa, whereas O$_3$ mixing ratios are reported at 1.0 hPa. Spherical projections are centered on 40$\degree$ N latitude and 225$\degree$ longitude. Red cross denotes the substellar point.} 
\end{center}
\end{figure*}

The presence of water vapor (H$_2$O$_v$), a canonical habitability indicator, could signify an active hydrological cycle. Photochemical and photolytic byproducts of H$_2$O$_v$ such as stratospheric OH and thermospheric H are produced by flare-initiated ion chemistry chains \citep{SolomonEt1981}. In our simulations, we find that hydrogen oxide family constituents are particularly sensitive to magnetic field assumptions due to their short lifetimes. Different magnetic field deflection geometries and host star UV spectral energy distributions contribute to different OH mixing ratio distributions (Figure 2e-h and 3e-h). During large stellar flares, stratospheric and mesospheric polar OH mixing ratios in the magnetized G-star planet simulation are ${\sim}10^{-8}$, but two orders of magnitude greater on the dayside of the unmagnetized K-star planet (${\sim}10^{-6}$).

The existence and persistence of stratospheric ozone in planetary atmospheres is fundamentally important for the protection and development of surface life \citep{Segura2018Handbook}. Simulated K- and M-star planet atmospheres (Figure 3a-l and Extended Data Figure 4a-l) experience greater instantaneous ozone destruction compared to magnetized G-star planets (Figure 2a-l). In the latter scenario with an Earth-similar magnetosphere, protons are funneled to the polar regions by magnetic field lines, whereas protons directly interact with the dayside atmospheres of the K- and M-star planets, enhancing ozone destruction. These differences are further compounded by redirection of protons to the polar nightside of G-star planets, initiating more sluggish chemical reactions than those that occur in the substellar regions of the unmagnetized planets. Ozone distribution differences between unmagnetized/weakly magnetized K- and M-star planets are due to the more active TRAPPIST-1 spectrum used (compared to that of HD85512), rapid horizontal mixing of chemical species, and strengthened downward transport on M-star planets.

To assess the role of flare frequency on ozone retention, we simulate three flare frequency assumptions (Supplementary Table 1 Exp. 8, 10 and 11; Extended Data Figure 3). For stellar activity approaching that of an optically-inactive M-dwarf (i.e., with cumulative flare index $\alpha = 0.7$; MUSCLES sample, \citep{LoydEt2018ApJa}), the computed total ozone column (see Methods) of the M-star planet gradually transitions to a depleted regime and establishes a new chemical steady-state (Extended Data Figure 6). For stars with activity levels similar to Proxima Centauri and AD Leo ($\alpha = 0.54$), we find that the ozone column experiences abrupt destruction from ${\sim}300$ Dobson Units (DU) to ${\sim}106$ DU over ${\sim}200$ Earth-days due to rapid erosion by incident stellar protons (Extended Data Figure 6). Our results could thus be used by observers to tie a measured flare frequency to cumulative effects on a planet’s atmosphere. 

\begin{figure*}[t] 
\begin{center}
\includegraphics[width=1\columnwidth]{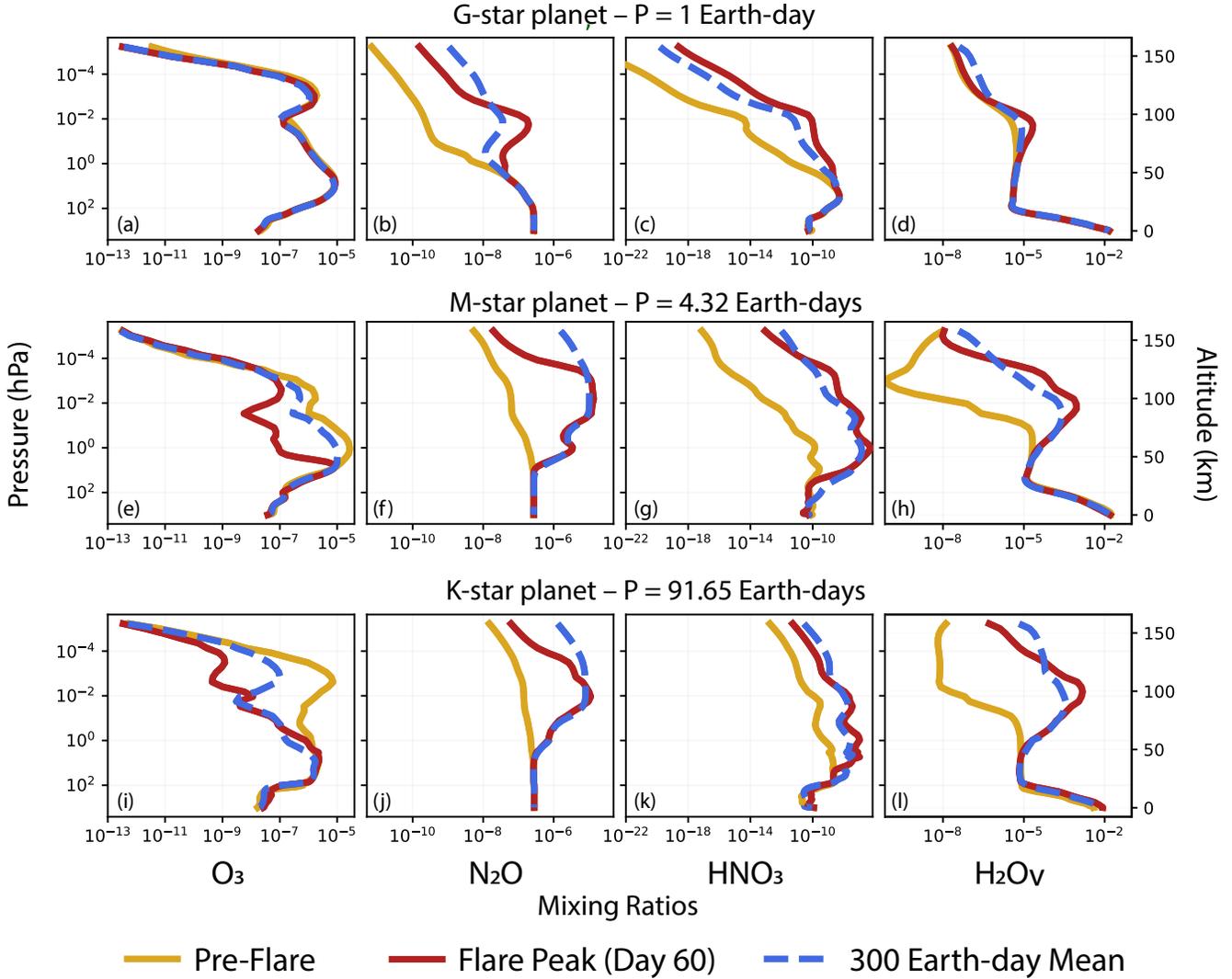}
\caption{\label{fig:profiles} {\bf Global mean vertical profiles of atmospheric species.} Simulated mixing ratios of O$_3$, N$_2$O, HNO$_3$, and H$_2$O$_v$ as a function of pressure and altitude during an initial steady state, the peak of a large flare, and averaged over a 300 Earth-day period. Conditions are shown for an Earth-similar planet around a G-star (a-d), K-star (e-h), and M-star (i-l) that experience flares with proton fluences of ${\sim}10^{14.5}$cm$^{-2}$.}
\end{center}
\end{figure*}

In addition to modeled flares, we use observed flares from the first {\it TESS} data release, specifically M-dwarfs TIC 671393 and 1636399 over 20-30 days of observation time (Extended Data Figure 2; Supplementary Table 1 Exp. 13 and 14). We find that stellar flares from the {\it TESS} data lead to more subtle changes in the chemical composition of the attendant planets, in comparison with the MUSCLES-based results that use modeled flare lightcurves with extrapolations to higher energies. For instance, simulated mesospheric ozone is halved at the end of the TIC 671393-based simulation (Extended Data Figure 5), whereas the full 300-day simulation using MUSCLES data results in a 1 to 2 order of magnitude decrease (Figure 3). The absence of repeated energetic flares and the short time-frame of the observed TESS data drive lower production rates of NO and UV photolysis of ozone. This conclusion is consistent with analysis of {\it TESS} across ${\sim}24,000$ samples, as very few stars exhibit continuous flares that exceed the $10^{34}$ erg threshold for ozone depletion \citep{GuntherEt2020AJ}.

Apart from stellar characteristics, flare influences are also controlled by the interplay between planetary properties, for example, between atmospheric mixing and photochemistry. Fast rotation, ($P < 6$ days) as in the case of our modeled G-star planet scenario, induces standing tropical Rossby waves  that disrupt meridional overturning circulation,  as opposed to extratropical Rossby waves in the case of the M-star planets and weak planetary waves in the case of the K-star planet. Without rapidly-rotating Earth-like planet deep wave breaking mechanisms or momentum injections into the stratosphere, slow-rotator stratospheric winds are effectively damped and prevent divergent meridional flow and planet-wide chemical transport, leading to confinement of flare induced species in the equatorial regions (Extended Data Figure 7; \citep{CaroneEt2018MNRAS}). However, fast rotation facilitates downward transport of ozone depleting agents such as NO$_{\rm x}$ into the mid-lower stratosphere (Extended Data Figure 8a). Conversely, slow rotation (i.e., K-star planet) allows NO$_{\rm x}$ to remain in the mesosphere/thermosphere (Extended Data Figure 8b). The simulated descent of flare-induced species is analogous to the advection of Earth’s NO$_{\rm x}$-rich airmasses into the stratosphere, driven by the stratospheric polar vortex and large-scale eddies  \citep{funke2005downward}. Thus, while the presence or absence of a planetary magnetic field plays a key role in governing ozone destruction (as polar ozone could be replenished by efficient meridional circulation), our results indicate that slow rotation (i.e., $P > 25$ days for an Earth-sized planet) can help maintain a stable global ozone layer against proton-initiated removal.

The elevated nitrogen and hydrogen oxides discussed above influence the formation and lifetimes of other atmospheric species such as N$_2$O, CH$_4$, HNO$_3$, and H$_2$O$_v$, especially for the K- and M-star planet scenarios. Effects on the G-star planet are generally less persistent as seen by lesser 300-day mean deviation from the pre-flare baselines (Figure 4a-d). For all scenarios, we find that flaring produces the largest magnitude alteration in nitrous oxide (N$_2$O), a biosignature (Figure 4; see Methods). Both HNO$_3$ and H$_2$O$_v$ mixing ratios are enhanced on average by two-to-three orders of magnitude and the enhancements are maintained with repeated flaring in the K- and M-star planet scenarios (Figure 4b-c). In contrast, CH$_4$ experiences stronger removal via reaction with ion-derived OH during flaring, leading to lower temporal-mean CH$_4$ mixing ratios (not shown). These results suggest that while biosignatures such as CH$_4$ are vulnerable to destruction during periods of strong flaring, bio-indicating ``beacon of life” species \citep{AirapetianEt2017Sci} could be prominently highlighted.

\begin{figure*}[t] 
\begin{center}
\includegraphics[width=0.85\columnwidth]{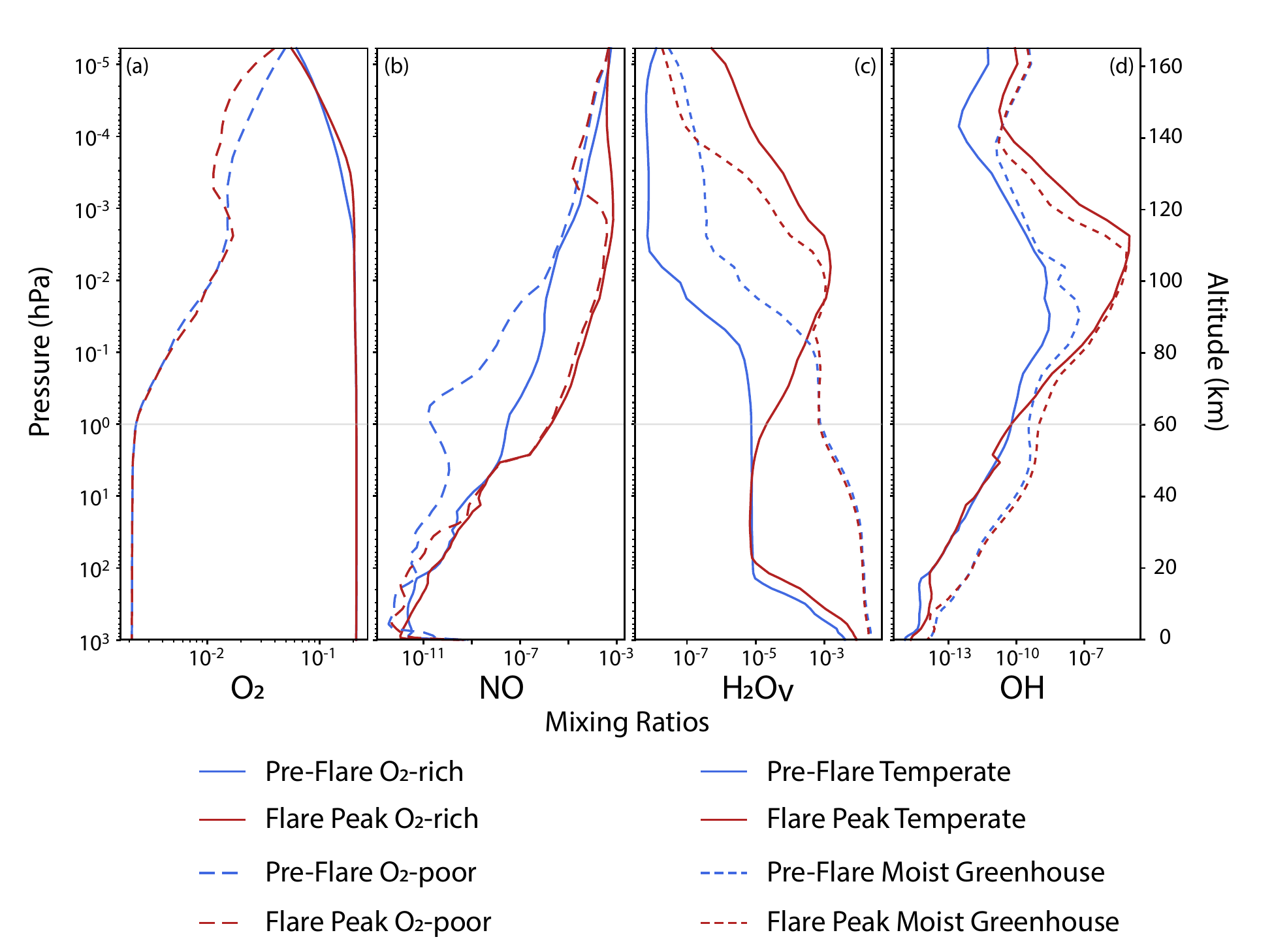}
\caption{\label{fig:rvp_profiles} {\bf Global mean vertical profiles of atmospheric species for non-modern-Earth climate archetypes.}  Simulated mixing ratios of O$_2$ (a), NO (b), H$_2$O$_v$ (c), and OH (d) under different climatic conditions. O$_2$ and NO are reported for simulations with O$_2$-rich and O$_2$-poor initial conditions. H$_2$O$_v$ and OH are shown for two different incident stellar fluxes: temperate (stellar flux = 1.0 $F_\oplus$) and moist greenhouse (stellar flux = 1.9 $F_\oplus$). The gray dashed lines denote the approximate transition from the lower atmosphere (stratosphere + troposphere) to the mesosphere and lower thermosphere region.  } 
\end{center}
\end{figure*}

Finally, we discuss flare-modulated atmospheric sensitivities of K-dwarf planet non-modern-Earth-similar compositions, including, O$_2$-poor Proterozoic-like initial conditions and a moist greenhouse state (Supplementary Table 1 Exp. 5 , and 6) -- common hypothetical exoplanetary conditions. These simulations show important departures from the modern-Earth composition baseline (Supplementary Table 1 Exp. 4). Interestingly, although the Proterozoic-like simulation contains orders of magnitude lower NO concentrations than that of the Earth-like baseline simulation during the pre-flare state, the flare peak NO concentration of the Proterozoic-like simulation approaches that of the latter (Figure 5b). This indicates that flare-induced NO could serve as potential proxy that are not directly detectable or challenging to observe (e.g., O$_2$). Repeated flaring in the moist greenhouse (specific humidity $Q_{{\rm H}2{\rm O}}  > 10^{-3}$) simulation leads to greater OH production and more rapid ozone destruction due to higher humidity than the Earth-like baseline simulation. In addition, in both the temperate and moist greenhouse simulations, the upper atmosphere is pushed into a classical moist greenhouse state \citep{Kasting1988Icarus}, despite the temperate simulation having relatively dry surface conditions (Figure 5c). This suggests that recurring flares via proton events could drive enhanced water loss through diffusion-limited escape even for planets that do not reside at the inner edge of the habitable zone. These putative non-Earth-archetypes demonstrate that flare-driven accumulation of nitrogen and hydrogen oxides could be reliable indicators of an N$_2$-O$_2$-H$_2$O-dominant atmosphere.

\begin{figure*}[t] 
\begin{center}
\includegraphics[width=1\columnwidth]{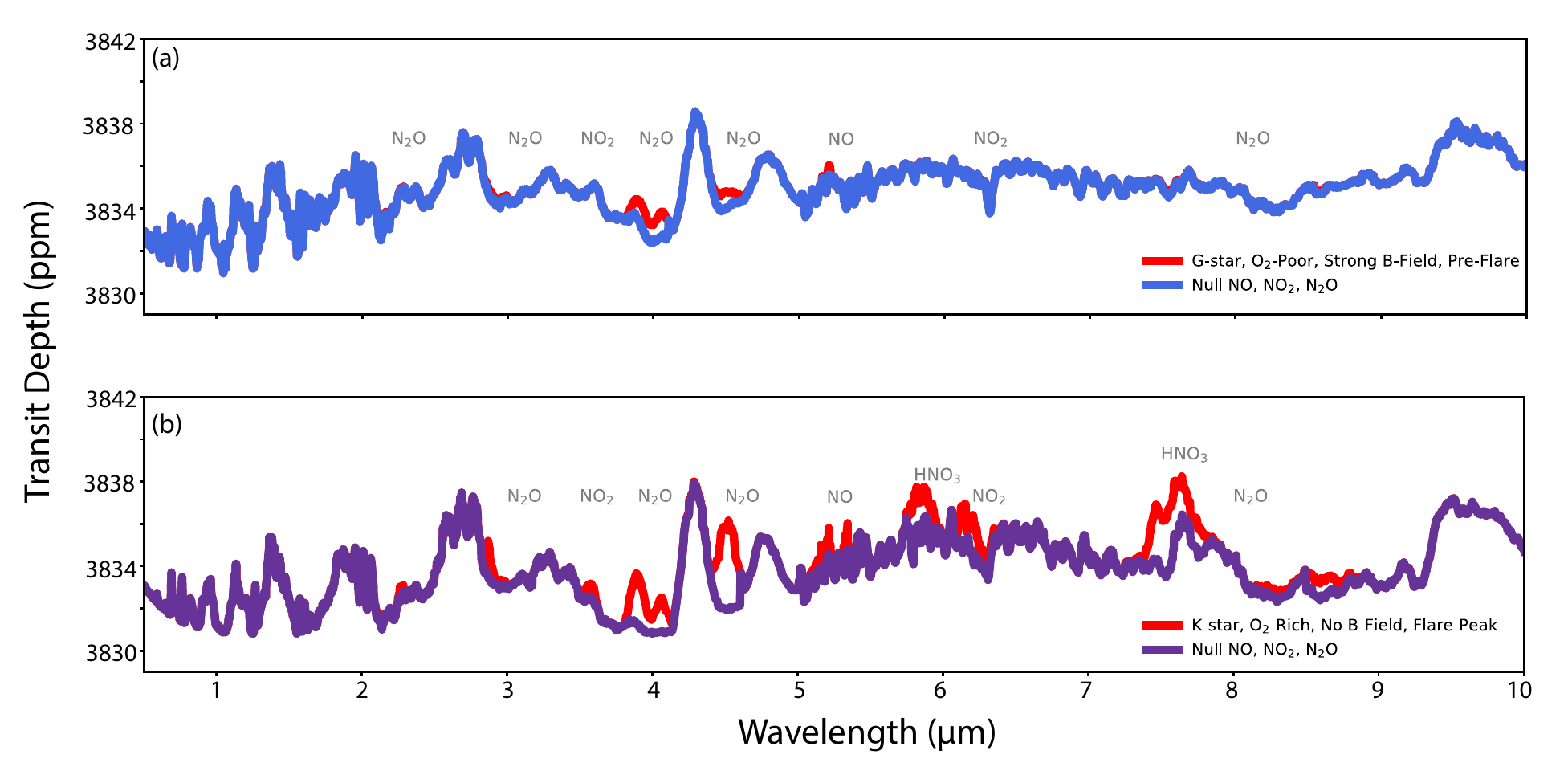}
\caption{\label{fig:specs} {\bf Simulated transmission spectra for two end member planetary scenarios.}  Modeled transit depth as a function of wavelength for simulated atmospheres of a magnetized Earth-like planet around a quiescent G-star without flare activity (a) and an unmagnetized synchronously-rotating planet around an actively flaring K-star (HD85512) (b). We assess the detectability of nitric oxide (NO), nitrous oxide (N$_2$O), nitrogen dioxide (NO$_2$), and nitric acid (HNO$_3$). The red curves contain variable amounts of simulated nitrogen oxide species, whereas the blue and purple curves contain no nitrogen oxides.   } 
\end{center}
\end{figure*}  

\section*{Observational Prospects and Implications (Discussion)}
Planetary transmission spectra, using our  chemistry-climate model outputs, demonstrate that stellar flaring induces spectral features of habitability indicators and biosignatures (Figure 6). Here we assess the detectability of nitrogen compounds for two endmember atmospheric scenarios from our suite of CCM simulations. Specifically, we compare the transit signals of NO, N$_2$O, NO$_2$, and HNO$_3$ on an O$_2$-poor magnetized planet orbiting a Sun-like star against those on an O$_2$-rich unmagnetized planet orbiting a K-dwarf. We find peak absorption depths of 2, 4, 3, and 6 ppm for the respective species in the latter scenario (Figure 6b). Despite transit depth shifts occurring above the cold trap and thus not muted by clouds \citep{FauchezEt2019ApJ,KomacekEt2020ApJL}, differences between pre-flare and flare peak features are less than the predicted noise floor of the James Webb Space Telescope (10-30 ppm; \citep{SuissaEt2020ApJ}). Moreover,  partial overlap of NO and NO$_2$ features with those of CO$_2$ and H$_2$O$_v$ at 4.3 and 5.5 $\mu$m obscures their signals. As such, detecting flare-driven biosignature fingerprints on synchronously rotating nitrogen-dominated Earth-sized exoplanets should await the development of larger telescopes with greater observing power and better instrument noise-floor control (i.e., with the noise-floor pushed to the ${\sim}1-2$ ppm level).

Other simulated spectral features, such as OH and O$_2$($^1$D), are likely only observable during or soon after a large flare, or in a system with a rapid succession of flares during transit measurements. This is due to the species' short chemical lifetimes, relaxation timescales, and rapid zonal mixing on non-synchronously-rotating G-star planets. Transmission features of biosignatures such as CH$_4$ and O$_3$ are predicted to be drastically reduced, as they react strongly with nitrogen and hydrogen oxides \citep{TabatabaEt2016A&A}. Note that these transient features arise primarily from species abundance changes in the mesosphere and lower thermosphere, and not from the stratosphere or troposphere (Figure 4). This finding is a result of our proton energy spectrum assumption (see Methods). Use of different proton energy spectrum assumptions could alter particle deposition depth, whole-column species abundances, and detectability.

Sudden increases in X-ray and EUV irradiation ($1.0  < \lambda < 100$ nm) -- which can energize, ionize, and dehydrate the upper atmosphere \citep{DongEt2017ApJLb,MordasiniEt2020A&A} -- are also associated with CMEs and stellar superflares \citep{DavenportEt2016ApJ,YangEt2017ApJ}. Thus, planets around active M-dwarfs may quickly lose their major high mean molecular weight species, while initially volatile rich atmospheres around less active K-dwarfs may be able to survive on geologic timescales \citep{Kite+Barnett2020PNAS}.

Here, we find that the convolved effects of magnetic field strength, radiation environment, and atmospheric circulation lead to substantial time-averaged (over ${\sim}1$ Earth year) chemical perturbations on flare-modulated K- and M-star planets. This result underscores the importance of constraining the temporal evolution of the host star spectra and luminosity to assess exoplanetary habitability. While we report the 3D effects of stellar flares on oxidizing atmospheres, strong flares could have other unexpected impacts on atmospheres with reducing conditions. For instance, hydrogen oxide species derived from stellar flares could destroy key anoxic biosignatures such as methane, dimethyl sulfide, and carbonyl sulfide \citep{Domagal-GoldmanEt2011AsBio}, thereby suppressing their spectroscopic features. However, new ionization rate profiles derived from a prognostic ion chemistry model will be needed to conduct analogous studies in atmospheric compositions dissimilar to Earth's. More speculatively, proton events during hyperflares may reveal the existence of planetary scale magnetic fields by highlighting particular regions of the planet (e.g., the poles; Figure 2a-d). By identifying nitrogen or hydrogen oxide emitting flux fingerprints during magnetic storms and/or auroral precipitation events, one may be able to determine the geometric extent of exoplanetary magnetospheres.

\section*{Methods}
\label{sec:method}

The U.S. National Center for Atmospheric Research (NCAR) Whole Atmosphere Community Climate Model (WACCM) was employed to simulate planetary atmospheres. Synthetic and observed flare timeseries and UV spectra are used as inputs to the climate model. Atmospheric transmission and emission spectra are computed using a radiative transfer model with updated molecular line-lists.

\subsection*{Chemistry-Climate Model} 

To simulate planetary atmospheres, we employ WACCM \citep{neale2010description,MarshEt2013JGR}, a high-top version of the Community Earth System Model v1.2, developed by NCAR. The model solves the primitive equations of fluid dynamics and thermodynamics, and includes self-consistent coupling of dynamics, chemistry, radiation, and thermodynamics. We use the Community Atmosphere Model v4 (CAM4) with the following modules: Community Atmospheric Model Radiative Transfer (CAMRT) radiation scheme \citep{zhang2003modified}, the Zhang-McFarlane scheme for deep convection \citep{zhang1995sensitivity}, and the Hack scheme for shallow convection \citep{Hack1994JGR}. The chemistry model is version 3 of the Modules for Ozone and Related Chemical Tracers (MOZART) chemical transport model \citep{KinnisonEt2007JGR}, which includes neutral and ionic constituents linked by 217 reactions. All surface gas fluxes are fixed at 1850 pre-industrial values, as industrial emission (e.g., nitrogen oxides; \citep{MontgomeryEt2018}) can affect surface chemistry. Note the small differences between exoplanet chemical model predictions \citep{YatesEt2020MNRAS,ChenEt2018ApJL}, suggesting the need for further model comparison efforts.

The oceanic model is a 30-meter thermodynamic ``slab" model with heat diffusion but no advection. The Community Land Model (CLM) v4 is used to model modern Earth continental configurations including pre-industrial surface features (e.g., vegetation, land type, and albedo).  Excluding ocean dynamics is appropriate for this study as we are primarily interested in photolytic and ion chemistry, processes that primarily take place above the stratopause. Moreover, ocean heat transport on tidally locked worlds is minimized by the North-South oriented continental configuration \citep{CheclairEt2019ApJL,YangEt2019ApJ}, or by the presence of an equatorial super-continent \citep{SalazarEt2020ApJL}.  3D climate simulations with a coupled dynamic ocean model of Proxima Centauri b show that an Earth-like continental setup eliminates day-to-night ocean heat transport, and the resulting global climate states are little changed compared to that found using a shallow slab ocean aquaplanet configuration with no dynamic heat transport  \citep{DelGenioEt2019AsBio}.

We used the finite-volume dynamical core and horizontal resolutions of $1.9\degree \times 2.5\degree$ ($144 \times 96$ number of grid cells). The vertical domain extends from the surface at 1013.25 hPa to model-top at $5.1 \times 10^{-6}$ hPa (145 km) with 66 levels. The vertical resolution is between 0.5-2 km in the troposphere and stratosphere and ${\sim}0.5$ scale height above the stratosphere. All simulations are integrated for at least 30 Earth years. Each simulation with the inclusion of flares and stellar proton events (SPEs) are then branched off from the final year and run for an additional 300 Earth days. All simulations adopt a model timestep of 900 seconds (15 minutes). 

\subsection*{Modifications for Synchronously-Rotating Climate Simulations}

WACCM is configured to simulate synchronously-rotating planets trapped in 1:1 spin-orbit resonance. For each climate scenario around HD85512 (K6V; \citep{GrayEt2006AJ}) and TRAPPIST-1 (M8V; \citep{SchmidtEt2007AJ}), the planet is placed at an orbital separation where the total incident TOA stellar flux is equal to that received by present-day Earth. The planetary rotation period is then set according to the stellar mass and luminosity Kepler's 3rd law \citep{KopparapuEt2016ApJ}. In all simulations, the orbital parameters (obliquity, eccentricity, and precession) are set to zero and the planets have 1 Earth-radius and 1 Earth mass. Continental configuration, topography, and surface albedo assume values of present-day Earth. The substellar point is placed over 180$\degree$ longitude. Earth's quasi-biennial oscillation, which is an observed alternating forcing between easterly and westerly zonal winds, is turned off.

\subsection*{Atmospheric Compositions}

To gauge how different atmospheric compositions respond to stellar activity, we simulate three different scenarios: (i) An ``O$_2$-rich" atmospheric composition (primary focus of the main text), i.e., pre-industrial Earth with N$_2$ (78\% by volume), O$_2$ (21\%), CH$_4$ (0.701 ppmv), N$_2$O (0.273 ppmv), and CO$_2$ (288 ppmv). ii) a Proterozoic-like atmosphere with very low O$_2$ concentrations (1\% Present Atmospheric Level (PAL)) to test the abundance of flare-modulated oxide compounds, though the extent to which O$_2$ could influence a planet's climate is controversial \citep{PoulsenEt2015SCI,PayneEt2016JRG,wade2019simulating}, and (iii) a moist greenhouse atmospheric composition  in which the decrease of tropospheric lapse rate and the expansion of the moist convection zones lead to the displacement of the cold trap to higher altitudes. Emergent moist greenhouse conditions are produced following the methodology of previous work \citep{KopparapuEt2016ApJ}, where we incrementally increase the incident stellar flux just prior to the onset of the incipient runaway greenhouse phase \citep{WolfEt2019ApJ}. The moist greenhouse scenario receives 1.9 times the present Earth's insolation, uses the stellar spectral energy distribution of HD85512, and has global-mean stratospheric H$_2$O mixing ratio of $1.28 \times 10^{-3}$. In all four scenarios, H$_2$O$_v$ and O$_3$ are spatially and temporally variable but are initialized at pre-industrial Earth values.

\subsection*{Expanded Radiative Transfer Components}

In addition to WACCM's high model top, its radiative transfer scheme includes non-LTE processes allowing for a more realistic assessment of the effects of stellar flares. In the MLT region, radiative transfer and dynamics are based on the thermosphere-ionosphere-mesosphere electrodynamics (TIME) GCM \citep{RobleEt1994GRL}. Key processes included are: neutral and high-top ion chemistry (ion drag, auroral processes, and solar proton events) and their associated heating reactions. Molecular diffusion via gravitational separation of different molecular constituents \citep{BanksET1973} is an extension to the nominal diffusion parameterization in CAM4. Below 65 km (local minimum in shortwave heating and longwave cooling), WACCM retains CAM4’s radiation scheme. Above 65 km, WACCM expands upon both longwave (LW) and shortwave (SW) radiative parameterizations from those of CAM3 and CAM4 \citep{CollinsEt2006}. WACCM uses thermodynamic equilibrium (LTE) and non-LTE heating and cooling rates in the extreme ultraviolet (EUV) and infrared (IR) \citep{FomichevE1998JGR}. In the SW (0.05 nm to 100 $\mu$m; \citep{Lean2000GRL,Solomon+Qian2005JGR,WangEt2005ApJ}), radiative heating and cooling are sourced from photon absorption, as well as photolytic and photochemical reactions. 
The native broadband radiation model of CAM4 is employed and not the newly introduced IR absorption coefficients \citep{KopparapuEt2017ApJ}. Such treatment is appropriate at temperatures $150 < T < 340$ K, which is the primary regime of interest in this study. 

\subsection*{Planetary Magnetic Field Assumptions}

To test the influence of a gravitational field on the global incident charged particle distribution, we parameterize the presence of planetary-scale magnetic fields as follows:

{\it a)} Magnetized Scenario: Protons are injected at polar latitudes ($> 60\degree$) across all longitudes, as incident particles are guided by the magnetic field lines to higher latitudes. This means that both the day and nightside receive comparable proton fluences due to the deflection geometry. 

{\it b)} Weakly or Anomalously Magnetized Scenario: Protons are injected a three different areas, with the assumption that the magnetic field's direction fluctuates wildly and originates from several poles. i) between  30$\degree$ and 60$\degree$ latitude, 120$\degree$ and 240$\degree$ longitude, ii) between  -30$\degree$ and -60$\degree$ latitude, 120$\degree$ and 240$\degree$ longitude, and iii) between 30$\degree$ and -30$\degree$ latitude, 300$\degree$ and 60$\degree$ longitude.   

{\it c)} Unmagnetized Scenario: Protons are directly injected on the substellar hemisphere between 90$\degree$ and 270$\degree$ longitude. No magnetic field deflection occurs in this scenario. 

In all the above cases, the vertical distribution of ion pair production rate (i.e., the proton spectrum) is based on Earth observations by Michelson Interferometer for Passive Atmospheric Sounding (MIPAS) instrument and the Geostationary Operational Environmental Satellite (GOES)-11 during the October 31$^{\rm st}$ 2003 geomagnetic storm \citep{LopezEt2005JGR,jackman2008short}. Calculation of self-consistent particle energy spectra (e.g., \citep{TabatabaEt2016A&A,GrenfellEt2012AsBio}) and FUV emissions relationships will be conducted in future work.

\subsection*{Stellar Spectra}

Effects of stellar activity across three spectral types are investigated: G, K, and M. For G-star simulations we use an observed and reconstructed solar irradiance spectrum \citep{lean1995reconstruction}. The input spectrum version is fixed in the year 1850 and no observed irradiance cycle is included. For K- and M-stars, two spectral types that bracket the endmember range of low mass stars were used: i) TRAPPIST-1 (M8V; $T_{\rm eff} = 2511$ K) data from the Mega-MUSCLES survey (Wilson et al. in review) and ii) HD 85512 (K6V; $T_{\rm eff} = 4715$ K) stellar SED from the MUSCLES survey (https://archive.stsci.edu/prepds/muscles/; version 2.2; \citep{FranceEt2016ApJ,LoydEt2016ApJ,YoungbloodEt2016ApJ}). Both spectra are binned at 1 Angstrom resolution with negative-flux bins removed via iterative-averaging as  statistical noise in the low-signal regions and the subtracted background level can result in negative fluxes. Presence of statistical noise in the signal and subtracted background levels necessitates this approach \citep{FranceEt2016ApJ}. Both stellar spectra are constant in time with exception of the FUV and NUV spectra (110-320 nm), which varies with the occurrence of flares (see below). We investigate wavelengths greater than 110 nm, which captures the $Ly-\alpha$ lines and FUV/NUV ranges that affect photochemistry (with peak ionization in the upper stratosphere at 0.1 to 0.01 hPa). Shorter wavelengths are not included for the sake of computational simplicity. But as soft X-ray and EUV wavelengths have peak ionization rates in the thermosphere and the ionosphere, they are unlikely to substantially perturb our conclusions in the stratosphere and mesosphere. 
One photosphere-only PHOENIX model is used. It assumes stellar metallicities of [Fe/H] = 0.0, alpha- enhancements of [$\alpha$/M] = 0.0, surface gravity values log g = 4.5. This serves as a benchmark case for a lower-limit flux estimate in the UV regime \citep{HusserEt2013A&A}.  In addition, we have benchmarked our M-star
planet results against using the high energy TRAPPIST-1 spectrum reconstructed
by \citet{PeacockEt2019ApJ} and \citet{TurbetEt2020SSR}.

\subsection*{Time Dependent Stellar Activity}
While the described stellar SEDs provide steady state spectra of our fiducial stars, assessment of transient stellar emissions requires time-evolving spectra. To compute time-dependent flares, we utilize: i) an open source M-dwarf flare Python module based on a large-scale campaign to characterize stellar FUV evolution \citep{LoydEt2018ApJa}, and ii) observed stellar flares identified by a convolutional neural network from \textit{TESS} data \citep{FeinsteinEt2020arXiv}. Proton fluence calculations are derived from these flares \citep{YoungbloodEt2017ApJ}. To isolate the roles of changes in rotation period and magnetic field strength assumptions,  all CCM experiments described are subjected to the same stellar flare timeseries inputs (e.g., Figure 1n and 2n). Realistically however, the amplitude and frequency of stellar flares correlates with the effective temperature of the host star \citep{CandelaresiEt2014ApJ} and the orbital semi-major axis of the planet \citep{SeguraEt2010AsBio,GrenfellEt2012AsBio}. While we assume all flares are ``direct-hits" upon the attendant planet, future work should employ probabilistic CME impact models to better improve realism of input flare energy and impact frequency \citep{KayEt2019ApJL}.

\subsection*{The MUSCLES M-dwarf Flare Model}

A stochastic flare model based on the Measurements of the Ultraviolet Spectral Characteristics of Low-mass Exoplanetary Systems (MUSCLES) Treasury Survey V \citep{LoydEt2018ApJa} is used to generate UV lightcurves. The MUSCLES Survey (HST observing program 13650) characterized the radiation environment of low-mass stars including the X-ray, XUV, and FUV, and NUV fluxes \citep{FranceEt2016ApJ}. 
The flare model is based on observed data from two stellar populations \citep{LoydEt2018ApJa}: the MUSCLE M-dwarf sample and four active stars Proxima Centauri, AD Leo, EV Lac, and AU Mic.  Each flare lightcurve is represented by a box car function followed by an exponential decay, simplified from the complex observed behaviors (e.g., multiple, sustained peaks) of flares. This model is used to generate a series of flares with equivalent durations $\delta E$ drawn from a power-law distribution (typical of observed flares; \citep{HawleyEt2014ApJ}), where $\delta E$ is defined by: 
\begin{equation}
    \delta_E = \int{\frac{F_f - F_q}{F_q}dt}
\end{equation}       

\noindent where $F_f$ is the flare flux, $F_q$ is the quiescent flux, and $dt$ is the flare duration.

To get the flare electromagnetic spectra, quiescent UV spectra of each stellar SED are multiplied by the active-to-quiescent flux ratio given by the model (though we note that SPEs are found to be responsible for ${\sim}99\%$ of the flare effects; \citep{SeguraEt2010AsBio,TilleyEt2019AsBio}). All flares assume blackbody temperatures of 9000 K, consistent with estimated color temperatures of M-dwarf flares between 7700 and 14000 K \citep{KowalskiEt2013ApJS}, though M dwarf flares at much hotter temperatures have been observed \citep{LoydEt2018ApJb,FroningEt2019ApJL}.

The flare occurrence rate $\nu$ is given by \citep{LoydEt2018ApJa}:

\begin{equation}
    \nu = \mu\left({\frac{\delta_E}{\delta_{\rm ref}}}\right)^{-\alpha}
\end{equation}
\noindent where $\mu$ is the rate constant, $\delta_{\rm ref}$ is the reference equivalent duration value ($\delta_{\rm ref} = 1000$ s), and $\alpha$ is the power-law index. 
Three different values of $\alpha$, 0.82, 0.7, and 0.54 are tested. All synthetic flares are assumed to have equivalent durations between $10^6$ and $10^9$ seconds. Flares with equivalent durations of $10^4$ secs are close to the largest observed flare \citep{LoydEt2018ApJa}, so our flares are based on power-law extrapolations. Flares below total energy of $10^{30}$  ergs and equivalent durations of $10^6$ secs are omitted. The choice of equivalent duration range (i.e., between $10^6$ and $10^9$) also  reflects typical equivalent durations of flares observed in the {\it U} and {\it Kepler} bands \citep{HiltonEt2011,HawleyEt2014ApJ}.

\subsection*{{\it TESS} Flare Identification via Deep Learning}

Apart from utilizing the aforementioned stochastic flares, we use flares observed by the \textit{Transiting Exoplanet Survey Satellite} (\textit{TESS}), which is a five-year photometric survey covering ${\sim}80\%$ of the observable sky \citep{RickerEt14}. Due to the time resolution necessary for observing stellar flares, we chose to search ${\sim}100$ pre-selected light curves based on their effective temperatures (i.e., between 3000 and 4000 K). The 2-minute light curves are hosted on the Mikulski Archive for Space Telescopes (MAST) and were downloaded using the \texttt{lightkurve} package (https://docs.lightkurve.org/). The two objects shown are TIC 671393 and TIC 1636399 where TIC denotes the {\it TESS} Input Catalog IDs. TIC 671393 has stellar mass of $0.54~M_\odot$, stellar radius $0.548~R_\odot$, and effective temperature 3096 K. TIC 1636399 has stellar mass of $0.53~M_\odot$ stellar radius $0.537 R_\odot$, and effective temperature 3266 K. We identified a total of 61 flares for TIC 671393 over the 25 days of observation time and  21 flares for TIC 1636399 over 24 days.

To ensure completeness in flare energies, we follow the methods of previous work \citep{FeinsteinEt2020arXiv} for flare identification. They trained a convolutional neural network (CNN) on a previously created flare catalog  \citep{GuntherEt2020AJ} and use the CNN as a sliding-box detector, where the ``probability" of a data point is part of a flare increases to 1 when a flare is within the sliding-box. As such, we ``predict" where flares occur using these models. Flare amplitudes and equivalent durations were calculated by fitting a Gaussian rise and exponential decay profile on a local region of the light curve around the flare. Use of CCN is advantageous as it can identify both large ($\delta > 10^6$ s) and small flares ($\delta < 10^4$ s). While the former is the focus of this study, small flares have important cumulative effects over longer timescales. All code used is part of the open-source \texttt{stella} Python package \citep{FeinsteinEt2020JOSS}.

Previous multi-wavelength observations of stellar activity suggest that optical events can serve  as proxies for the initial heating of the chromosphere, as optical flares are found to precede the X-ray, EUV, and UV flares of the impulsive phase \citep{GudelEt2004A&A,StelzerEt2006A&A}. Thus while the \textit{TESS} data do not provide UV spectra, we assume that the UV flare frequencies over the course of $> 20$ days are qualitatively similar to those in the observed IR and optical. Simulations that use the \textit{TESS} data as inputs assume TRAPPIST-1 steady stellar SED from the Mega-MUSCLES survey (Wilson et al. in review).

\subsection*{Proton Fluences and Ionization Rates}

Incident charged particles of stellar origin are associated with large flares and CME-like events \citep{GopalswamyEt2012,SinnhuberEt2012,Atri2017MNRAS}. While direct observations of energetic particle emissions during CMEs are not available (i.e., only signatures of CMEs are observable; \citep{FranciosiniEt2001A&A}), we follow previous studies by using solar scalings based on near-Earth satellite data. We assume that all of the particles are protons. We compute the expected peak proton fluences from the SiIV energy of stellar flares \citep{YoungbloodEt2017ApJ}: 

\begin{equation}
    {\rm log}~F_{> 10 ~{\rm MeV}} = 1.20~{\rm log}F_{\rm SiIV} + 3.27
\end{equation}

\noindent where $F_{> 10 ~{\rm MeV}}$ is the proton fluence. The derived fluences all follow the M-dwarf flare model generated lightcurve shapes/durations. Since there is a linear relationship between the proton flux and production rate of ion pairs \citep{jackman2008short}, we scaled the input ionization rates comparing our estimated proton fluence with that of 2003 Halloween SPE (an order of magnitude lower than the Carrington event in 1710; \citep{jackman2008short,LopezEt2005JGR}). The ion pair production rates, provided in the  Solar Influence for SPARC (SOLARIS) website (https://solarisheppa.geomar.de/solarprotonfluxes; newly updated in March 2019) and derived from proton flux measurements by GOES 11 instrument, are then applied as daily averages during each flare peak. Daily cadences are appropriate in this pilot study as the cascading NO$_{\rm x}$ and HO$_{\rm x}$ reactions are much  faster \citep{EjzakEt2007ApJ} than the flare and model timesteps, but future work should employ higher temporal resolutions to better resolve stellar activity on hourly timescales (e.g., \citep{pettit2018effects}). 

Other methods to calculate ionization rates due to stellar flares or galactic-sourced cosmic rays e.g., the air-shower approach \citep{GrenfellEt2012AsBio}, have shown to compare well with the approach taken here and those in prior studies \citep{SeguraEt2010AsBio}. Note that the majority of stellar and exoplanetary studies uses the same peak size distribution  functions from solar events  \citep{BevlovEt2005,CliverEt2012ApJL}. However, large discrepancies exist between published peak size distributions due to the different underlying physical mechanisms driving these events   \citep{HerbstEt2019A&Aa}. Thus the conclusions established from photochemical models, even with the same flare inputs, would likely be contingent upon the specific function used.

\subsection*{Stellar UV and Proton Event Initiated Atmospheric Chemistry}

WACCM includes a range of chemical reactions necessary to fully account for the effects of stellar activity. Interaction of UV photons with trace gases typically lead to dissociation via photolysis \citep{Jacob1999}:

\begin{equation}
    {\rm H}_2{\rm O} + h\nu (175 < \lambda < 200 {\rm nm}) \rightarrow {\rm H} + {\rm OH}
\end{equation}

\begin{equation}
    {\rm O}_3 + h\nu (\lambda < 320 {\rm nm}) \rightarrow {\rm O}_2 + {\rm O}(^1{\rm D})
\end{equation}

Some important daytime(side) photochemical reactions are \citep{Ball2014ECG}:

\begin{equation}
     {\rm O}(^1{\rm D}) + {\rm H}_2{\rm O} \rightarrow 2{\rm OH}
\end{equation}

\begin{equation}
    {\rm HO}_2 + {\rm NO} \rightarrow {\rm H} + {\rm NO}_2
\end{equation}

\begin{equation}
    {\rm NO}_2 + h\nu (\lambda < 420 {\rm nm}) \rightarrow {\rm NO} +  {\rm O}(^3{\rm P}) 
\end{equation}

Particle precipitation due to SPEs also influence atmospheric chemistry. SPEs produce charged particles (protons and secondary electrons) that causes excitation and subsequent dissociation of ambient gaseous constituents. Ground state and excited state nitrogen are produced via \citep{JackmanEt2005JGR, PorterEt1976JCP}:

\begin{equation}
    {\rm N}_2 + {\rm e}^- \rightarrow {\rm 2N}(^4{\rm S}) + {\rm e}^- 
\end{equation}
or

\begin{equation}
    {\rm N}_2 + {\rm e}^- \rightarrow {\rm 2N}(^2{\rm D}) + {\rm e}^- 
\end{equation}

\noindent where e$^-$ represents secondary electrons produced by incident protons. We assume that 1.25 N atoms are produced per ion pair (specifically 0.55 N($^4$S) ground state atoms and 0.7 N($^2$D) excited state atoms; \citep{PorterEt1976JCP}). 

Excited and ground state nitrogen can subsequently produce NO via:

\begin{equation}
   {\rm N}(^2{\rm D}) + {\rm O}_2  \rightarrow {\rm NO} +  {\rm O}(^3{\rm P})
\end{equation}

\begin{equation}
     {\rm N}(^2{\rm S}) + {\rm O}_2  \rightarrow {\rm NO} +  {\rm O}(^3{\rm P})
\end{equation}

or remove NO via:

\begin{equation}
     {\rm N}(^2{\rm S}) + {\rm NO} \rightarrow {\rm N}_2 +  {\rm O}(^3{\rm P})
\end{equation}

Two HO$_{\rm x}$ species are produced per ion pair. Parameterization of prognostic ionic water cluster reaction networks is done via: 

\begin{equation}
    {\rm H}_2{\rm O} +  {\rm Ion}^+  \rightarrow {\rm H} + {\rm OH} + {\rm Ion}^+
\end{equation}

Increase HO$_{\rm x}$ species lead to catalytic ozone destruction in the stratosphere via \citep{SolomonEt1981,SinnhuberEt2012}:

\begin{equation}
     {\rm OH} + {\rm O}_3 \rightarrow {\rm HO}_2 + {\rm O}_2
\end{equation}

\begin{equation}
      {\rm HO}_2 + {\rm O} \rightarrow {\rm OH} + {\rm O}_2
\end{equation}

and the mesosphere via:

\begin{equation}
     {\rm H} + {\rm O}_3 \rightarrow {\rm OH} + {\rm O}_2
\end{equation}

\begin{equation}
     {\rm OH} + {\rm O} \rightarrow {\rm H} + {\rm O}_2
\end{equation}
While OH and H are act on on short timescales (${\sim}$ hours), the prolonged lifetimes of NO$_{\rm x}$ species can deplete stratospheric ozone via:

\begin{equation}
     {\rm NO} + {\rm O}_3 \rightarrow {\rm NO}_2 + {\rm O}_2
\end{equation}

\begin{equation}
    {\rm NO}_2 + {\rm O} \rightarrow {\rm NO} + {\rm O}_2
\end{equation}

Apart from contributing to ozone loss, enhanced NO and OH during large stellar flares can also modulate the abundance of other important biosignatures and habitability indicators (see e.g., \citep{ScheucherEt2018ApJ,HerbstEt2019A&Ab}). Example reactions are:

\begin{equation}
    {\rm N} + {\rm NO}_2 \rightarrow {\rm N}_2{\rm O} + {\rm O}
\end{equation}

\begin{equation}
    {\rm NO}_2 + {\rm OH} + {\rm M}  \rightarrow {\rm HNO}_3 + {\rm M}
\end{equation}

\begin{equation}
    {\rm CH}_4 + {\rm OH} \rightarrow {\rm CH}_3 + {\rm H}_2{\rm O}
\end{equation}

\noindent where NO$_2$ can be photochemically derived from the cascading N$_2$ dissociation initiated by SPEs, where OH can come from water vapor photolysis, ion chemistry, and reaction between H$_2$O + O($^1$D). N$_2$O can also be produced from the reaction between NO and NH.

\subsection*{Atmospheric Spectroscopy Model}

To translate WACCM output into observable predictions, we use the Simulated Exoplanet Atmosphere Spectra (SEAS) model (Zhan et al. in press) to compute atmospheric spectra. SEAS is a radiative transfer code that calculates the attenuation of photons by molecular absorption and Rayleigh/Mie scattering as the photons travel through a hypothetical exoplanet atmosphere. The simulation approach is similar to previous work \citep{KemptonEt2017ApJL,KemptonEt2009ApJ}. The molecular absorption cross-section for O$_2$, H$_2$O, CO$_2$, CH$_4$, O$_3$, and H are calculated using the HITRAN2016 molecular line-list database \citep{GordonEt2017}. The SEAS transmission spectra are validated through comparison of its simulated Earth transmission spectrum with that of real Earth counterparts measured by the Atmospheric Chemistry Experiment (ACE) data set \citep{BernathEt2015GRL}. For more details on SEAS, please see Section 3.4 in Zhan et al., (in press).

\subsection*{Day-Night Mixing Ratio \& Ozone Column}

Mixing ratio contrasts of various gases between the day and nightside are defined as \citep{ChenEt2018ApJL}: 

\begin{equation}
    C_{\rm diff} = \frac{C_{\rm day} - C_{\rm night}}{C_{\rm globe}} 
\end{equation}

\noindent where $C_{\rm day}$ is the dayside hemispheric mixing ratio mean, $C_{\rm night}$ the nightside mean, and $C_{\rm globe}$ the global mean. We compute $C_{\rm diff}$ by vertically-averaging the mixing ratios between 1 and $1 \times 10^{-4}$ hPa. 

The column number density of species $i$ is defined as:

\begin{equation}
    N_i  = \int n_i dz 
\end{equation}

\noindent where $n_i$ is the volume number density of species $i$ and $z$ is the vertical coordinate. Here, we compute the total ozone column  by summing up the volume number density in all 66 levels. The number density of ozone in each level is calculated via:

\begin{equation}
    n_{\rm O3} = C_{\rm O3} \frac{N_A P_i}{R T_i}
\end{equation}

\noindent where $C_{\rm O3}$ is the mixing ratio of ozone, $N_A$ is the Avogadro's number, $R$ is the gas constant (8.31 J mol$^{-1}$ K$^{-1}$), and $P_i$ and $T_i$ are the atmospheric pressure and temperature at level $i$. The total ozone column is typically expressed in Dobson Units (DU). 1 DU = $2.686 \times 10^{20}$ m$^{-2}$ or $2.686 \times 10^{16}$ cm$^{-2}$.

\subsection*{Data Availability}
The data that support the plots within this paper and other findings of this study 
are available from the corresponding author upon request. The raw 
data are publicly available at \\https://mast.stsci.edu/portal/Mashup/Clients/Mast/Portal.html (MAST) and \\https://archive.stsci.edu/prepds/muscles/ (MUSCLES). and the solar ion pair production rates are available at https://solarisheppa.geomar.de/solarprotonfluxes.

\subsection*{Code Availability}
The unmodified climate model used in this study is available for public download at\\ http://www.cesm.ucar.edu/models/cesm1.2/cesm/doc/usersguide/x290.html. Components of the modified version of the climate can be obtained via the \texttt{ExoCAM},\\ https://github.com/storyofthewolf/ExoCAM and by request from ETW. The \texttt{stella} package can be downloaded at \\https://github.com/afeinstein20/stella.
The remaining codes that support the results within this paper and other findings of this study are available from the authors on request.

\subsection*{Acknowledgements}

HC and DEH acknowledge support from the Future Investigators in NASA Earth and Space Science and Technology (FINESST) Graduate Research Award 80NSSC19K1523. HC thanks Parke Loyd for assistance in the use of his MUSCLES flare code and for sharing it with the public. ZZ acknowledges support from the MIT BOSE Fellow program, the Change Happens Foundation, and the Heising-Simons Foundation. ETW acknowledges support from NASA Habitable Worlds grant 80NSSC17K0257. ADF acknowledges support from NSF Graduate Research Fellowship Program grant DGE-1746045. We thank A. Gu for ozone variability analysis inspiration and the QUEST high performance computing facility at Northwestern University for computational and staff resources. Any opinions, findings, and conclusions or recommendations expressed in this material are those of the author(s) and do not necessarily reflect the views of the funding agencies. 

This paper includes data collected by the \textit{TESS} mission. Funding for the \textit{TESS} mission is provided by the NASA Explorer Program. \textit{TESS} data were obtained from the Mikulski Archive for Space Telescopes (MAST). STScI is operated by the Association of Universities for Research in Astronomy, Inc., under NASA contract NAS5-26555. Support for MAST is provided by the NASA Office of Space Science via grant NNX13AC07G and by other grants and contracts. 

\subsection*{Author Contributions}

HC, ETW, and DEH conceived and designed the study. HC conducted the numerical model simulations and data analysis. ZZ performed the radiative transfer model simulations. AY provided stellar input data from the MUSCLES and Mega-MUSCLES survey. ADF performed the machine learning \textit{TESS} data reductions. HC wrote the manuscript with input from all co-authors. 

\subsection*{Competing Financial Interests}

The authors declare no competing financial interests.

\renewcommand{\figurename}{Extended Data Fig.}
\setcounter{figure}{0}

\begin{figure*}[t] 
\begin{center}
\includegraphics[width=0.8\columnwidth]{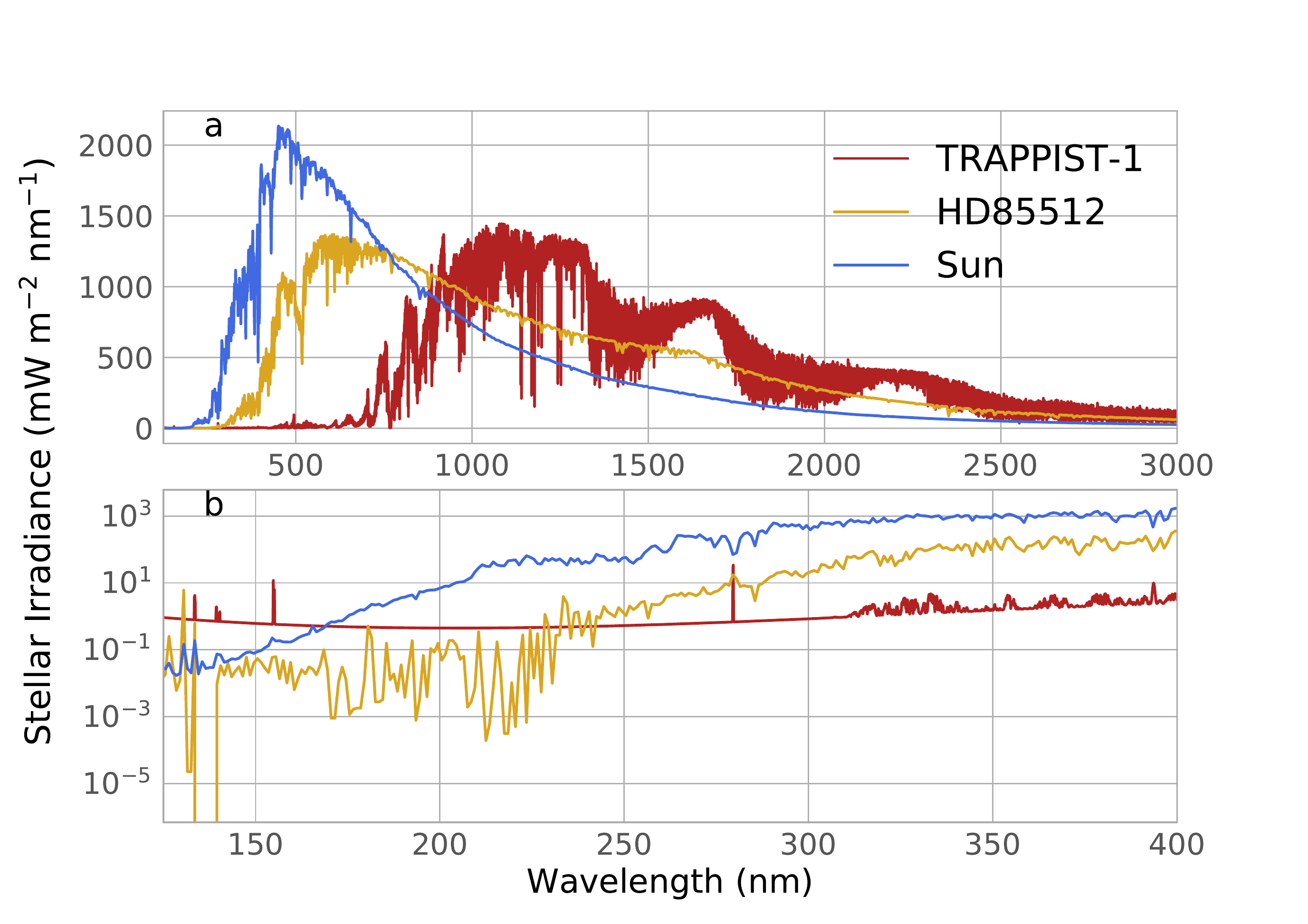}
\caption{\label{fig:stellar_spec_supp}  Input broadband (a) and UV (b) spectral energy distributions  for the Sun, HD85512, and TRAPPIST-1. The Sun represents the G-star archetype, HD85512 a K-star, and TRAPPIST-1 a late M-star. We refer to the stellar spectral types these stars represent (G-star, K-star, and M-star) instead of the specific star in the main text and throughout the paper. } 
\end{center}
\end{figure*}

\begin{figure*}[t] 
\begin{center}
\includegraphics[width=0.8\columnwidth]{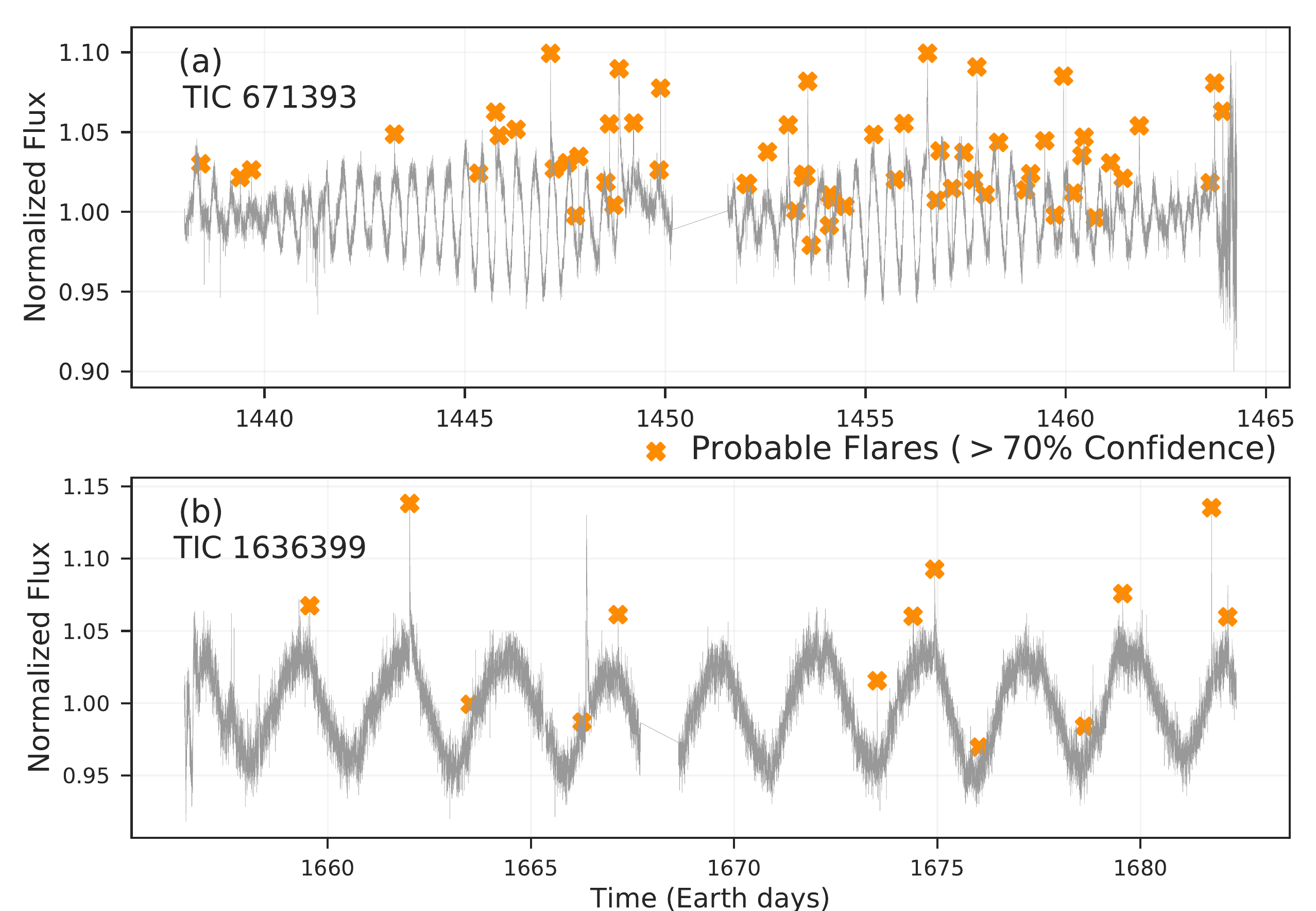}
\caption{\label{fig:tess_input} Timeseries of TESS lightcurves used in this study. The stellar data used are those of TIC 671393 (a) and TIC 1636399 (b), showing identified flares by orange ``$\times$"s. Flares are identified by a convolutional neural network algorithm described in Feinstein et al. (2020).} 
\end{center}
\end{figure*}

\begin{figure*}[t] 
\begin{center}
\includegraphics[width=0.8\columnwidth]{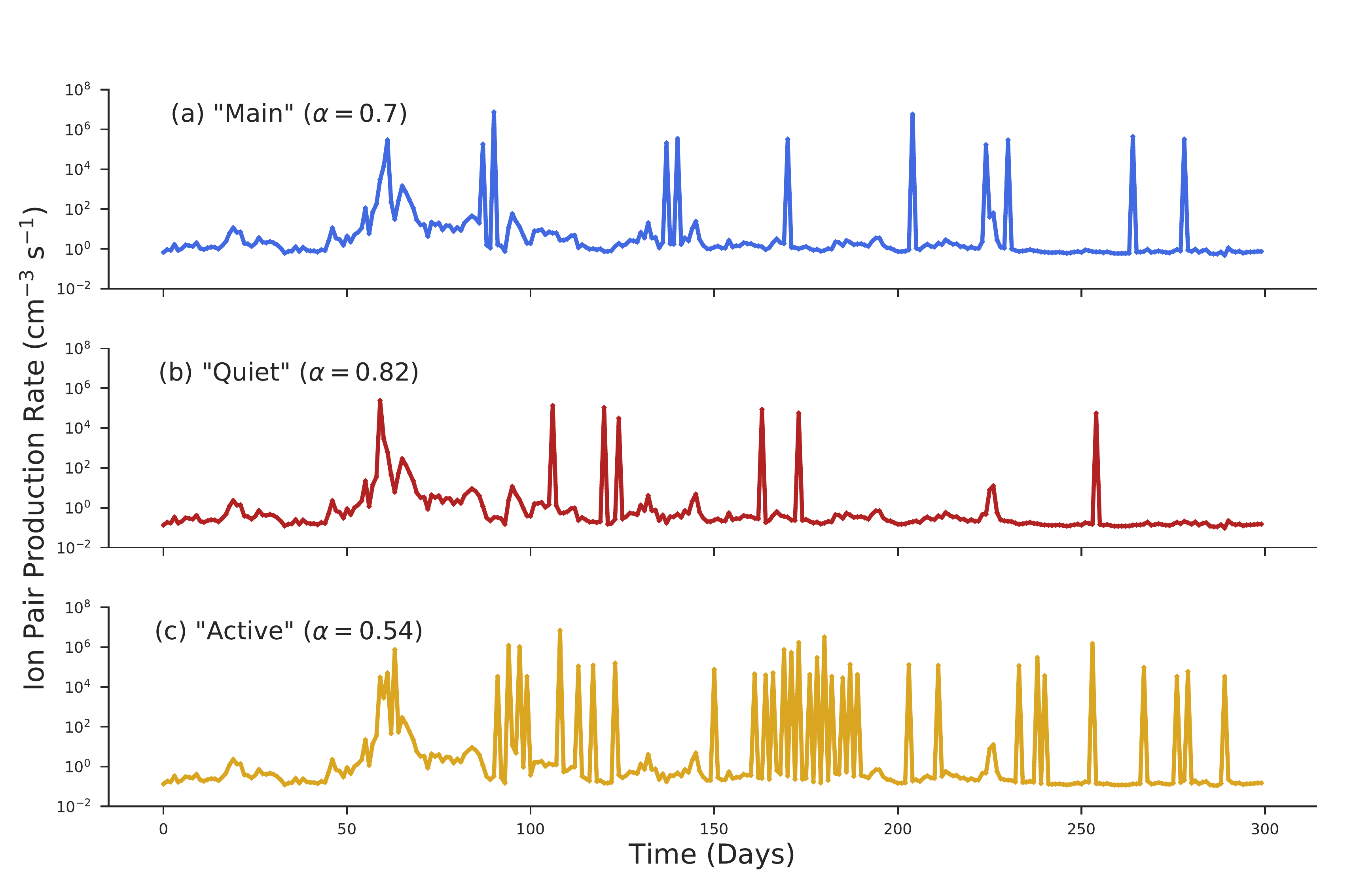}
\caption{\label{fig:ion_prod_supp}   Three scenarios of vertical-mean  ionization rates used as inputs to explore the effects of flare frequency. Three different assumptions are investigated: $\alpha = 0.7, 0.82, 0.54$.  Supplementary Table 1 lists the specific experiments and their assumed flare frequency.  } 
\end{center}
\end{figure*}

\begin{figure*}[t] 
\begin{center}
\includegraphics[width=0.8\columnwidth]{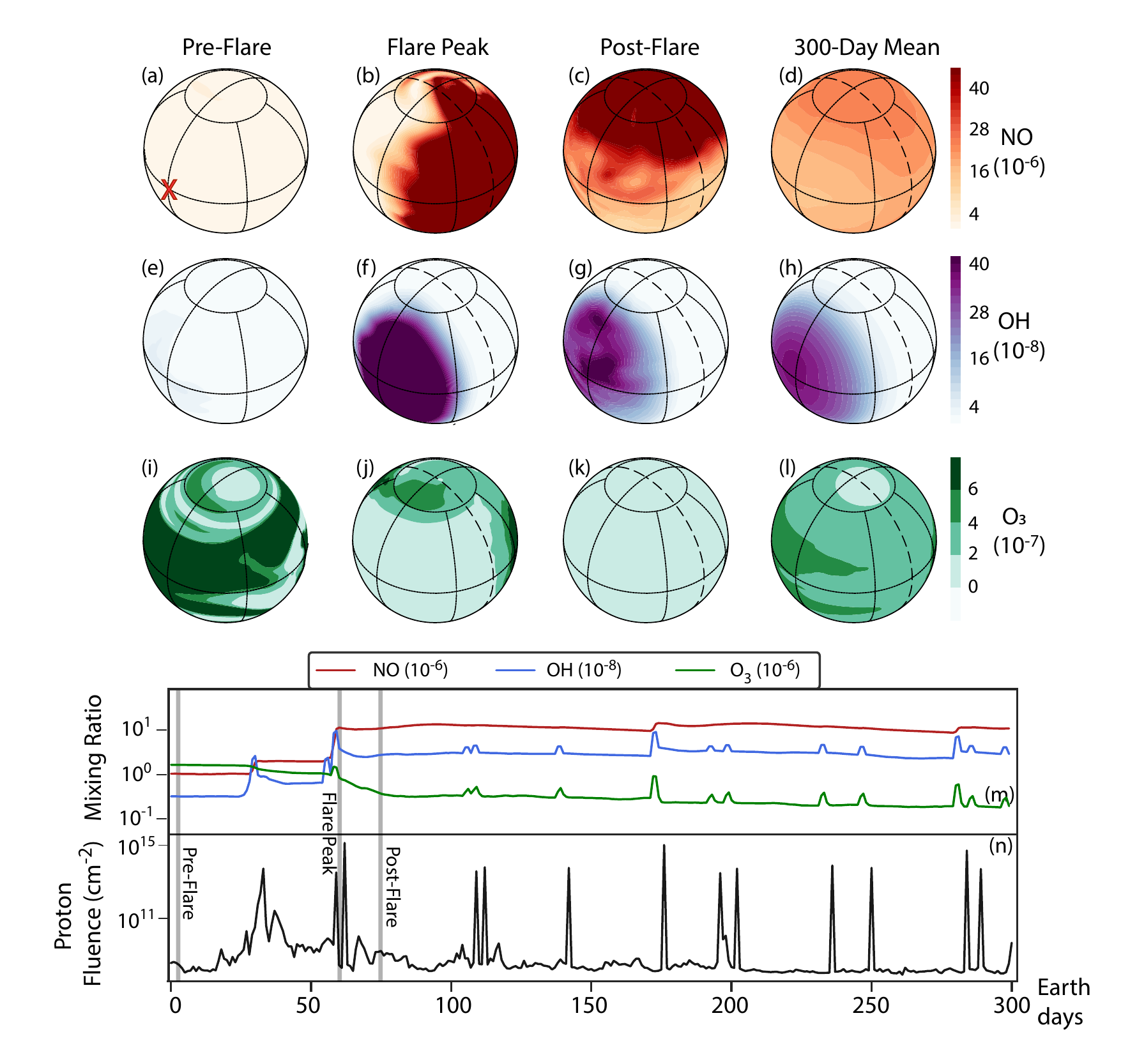}
\caption{\label{fig:trappist1}  Spatial and temporal atmospheric effects of repeated stellar flaring on an M-star planet. Simulated global time slice distributions of upper atmospheric NO (a-d), OH (e-h), and O$_3$ (i-l) concentrations and their global average time-series (m) that result from exposure to flares with time-evolving proton fluences (n). The simulated planet rotates around M-star TRAPPIST-1 synchronously and has a weak magnetic field.  and OH mixing ratios are reported at 0.1 hPa, whereas O$_3$ mixing ratios are reported at 1.0 hPa. Spherical projections are centered on 40$\degree$ N latitude and 225$\degree$ longitude. Red cross denotes the substellar point.} 
\end{center}
\end{figure*}

\begin{figure*}[t] 
\begin{center}
\includegraphics[width=0.9\columnwidth]{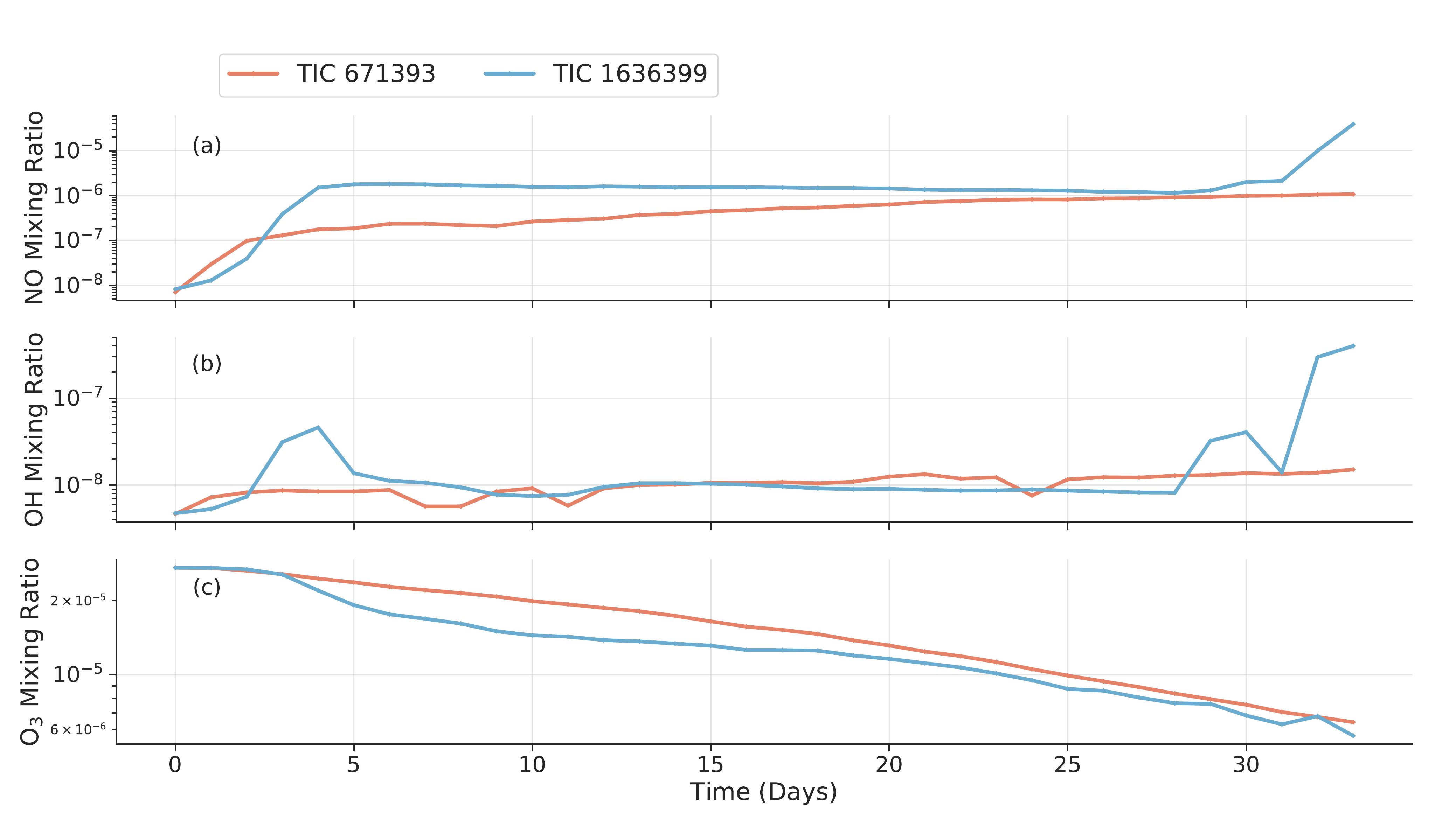}
\caption{\label{fig:tess_out}   Temporal evolution of global-mean mixing ratios of NO, OH, and O$_3$  experiencing {\it TESS} flares. Result demonstrate that small flares over a short timespan do not substantially affect exoplanetary atmospheres. NO and OH mixing ratios are reported at 0.1 hPa, whereas O$_3$ mixing ratios are reported at 1.0 hPa.
} 
\end{center}
\end{figure*}

\begin{figure*}[t] 
\begin{center}
\includegraphics[width=1.0\columnwidth]{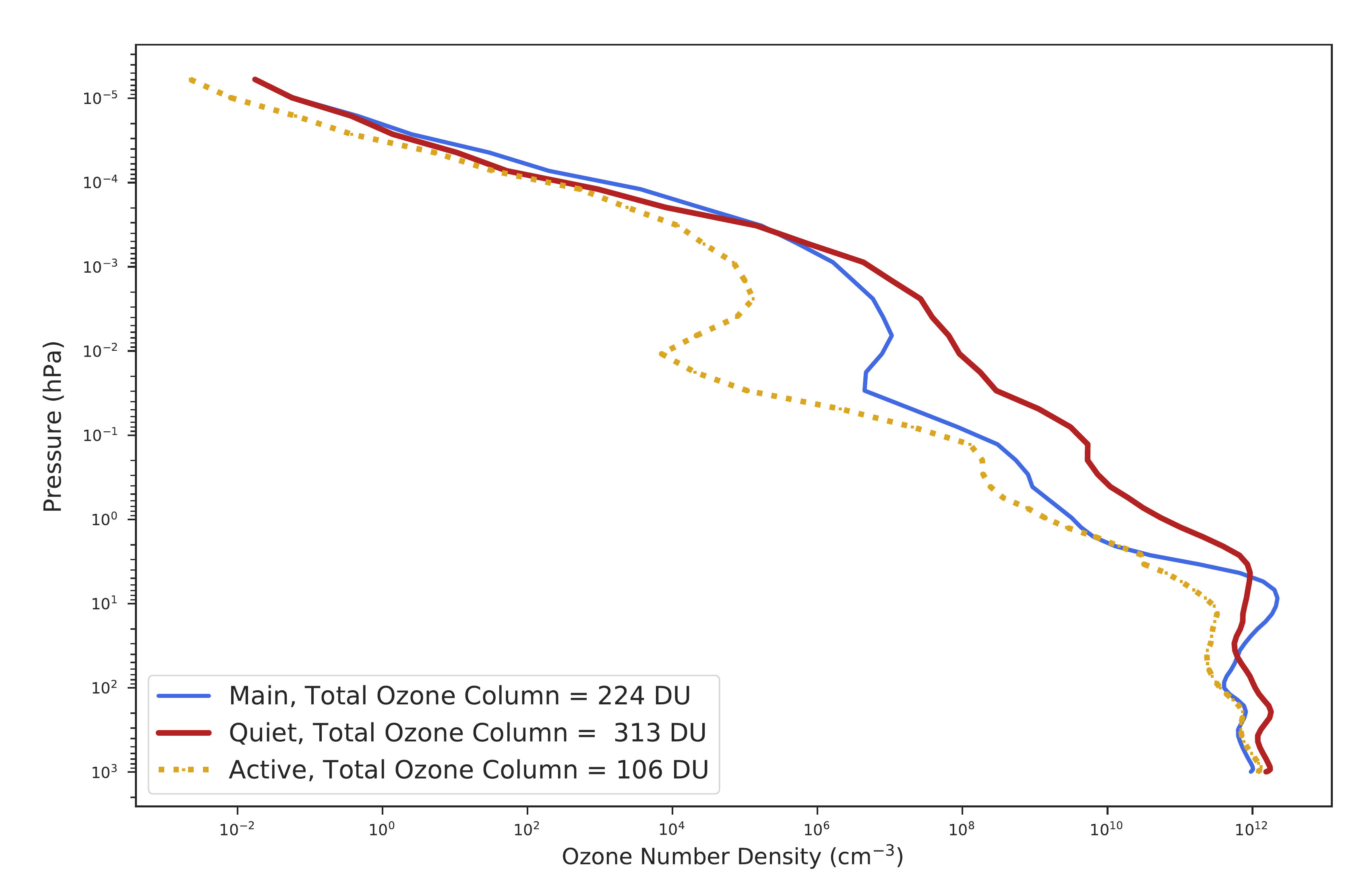}
\caption{\label{fig:oz_profiles}  Global-mean vertical profiles of ozone number density at three different stellar flare frequencies. These results show the cumulative effect (300 Earth days) of repeated stellar flares.  } 
\end{center}
\end{figure*}

\begin{figure*}[t] 
\begin{center}
\includegraphics[width=0.9\columnwidth]{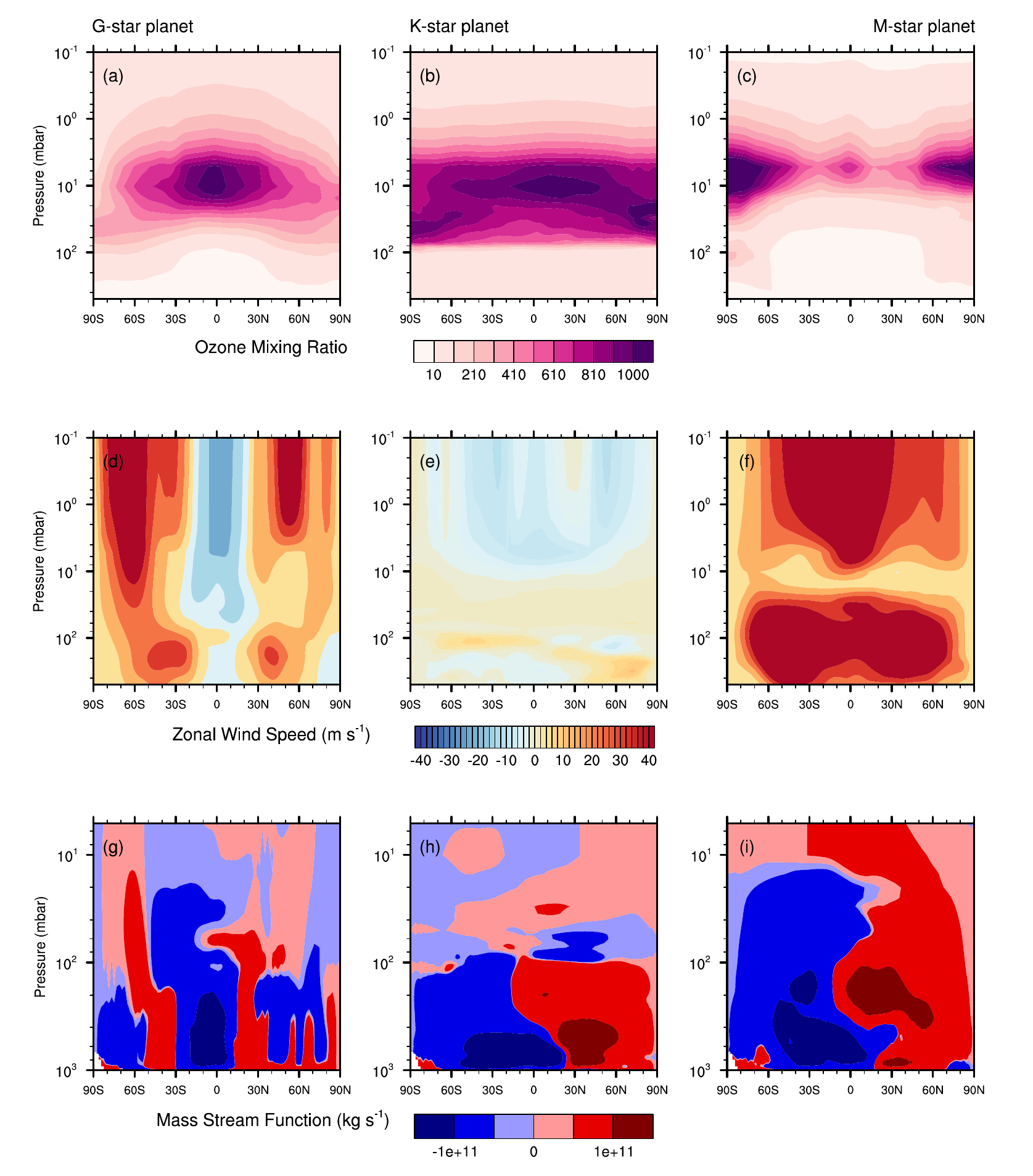}
\caption{\label{fig:oz_trans_supp} Zonal mean of zonal wind, O$_3$ mixing ratios, and meridional circulation stream functions for hypothetical O$_2$-rich planets around a G-star, K-star, and M-star as denoted. Results demonstrate the convolved effects of dynamics and atmospheric chemistry. } 
\end{center}
\end{figure*}

\begin{figure*}[t] 
\begin{center}
\includegraphics[width=0.7\columnwidth]{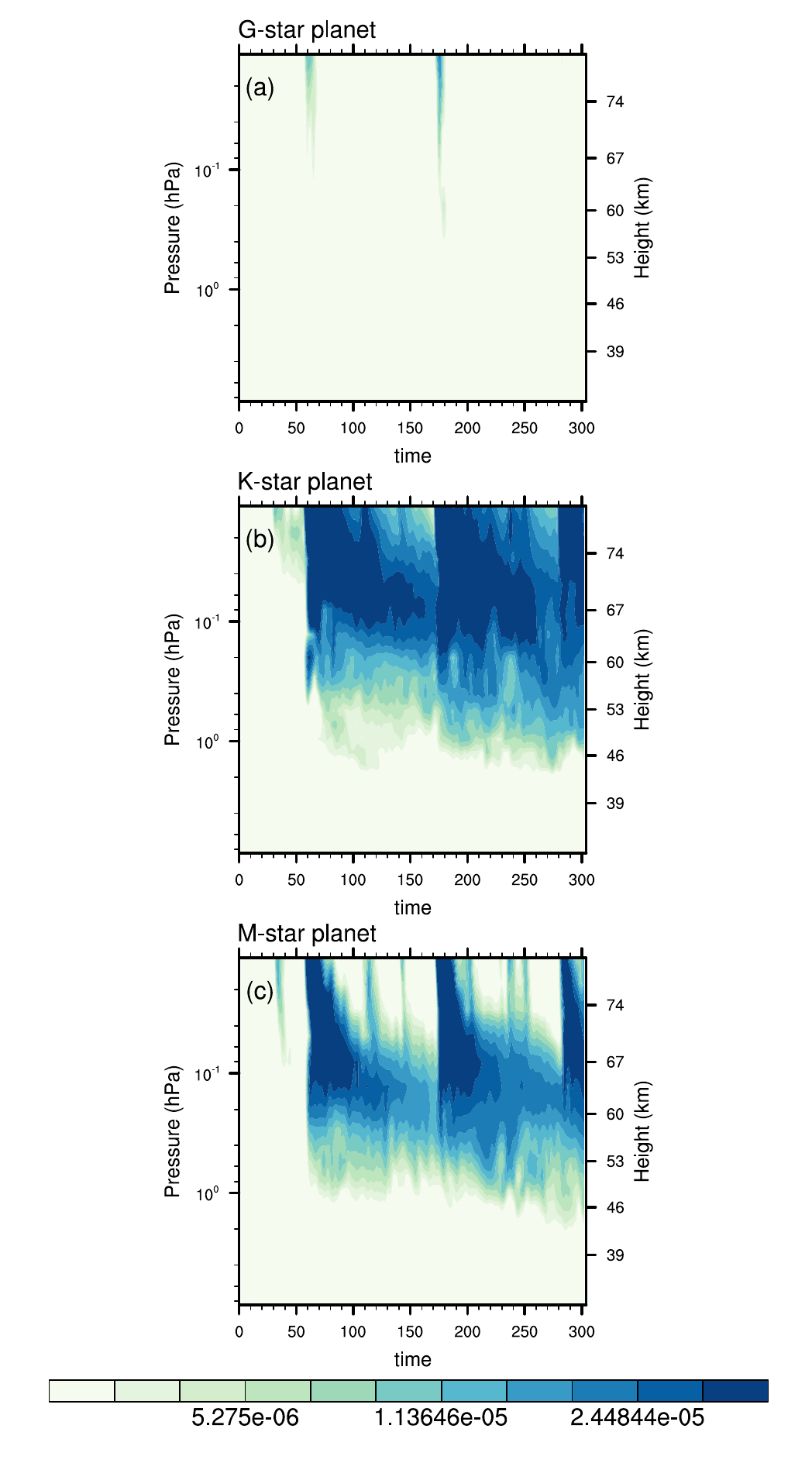}
\caption{\label{fig:no_timeseries_supp} NO concentration averaged over the poles ({{\rm latitude}}$ > 65\degree$) as a function of time and pressure for hypothetical O$_2$-rich planets. The rotation periods of these simulations are 24 hours, 92 Earth days, and 4.32 Earth days around a G-dwarf (a), K-dwarf (b), and M-dwarf (c) star. Note the log$_{\rm 10}$-scale. } 
\end{center}
\end{figure*}

\end{document}